\documentclass[11pt]{article}
\usepackage{times,graphicx,theorem,epsfig,lscape,dcolumn,multirow,epic,float}
\usepackage{amsmath,amssymb}
\usepackage[normalem]{ulem}
\usepackage{natbib}
\usepackage[figuresright]{rotating}
\usepackage[english]{babel}
\usepackage{booktabs}
\usepackage{hhline}
\usepackage{bm}
\usepackage{enumerate}
\usepackage{verbatim}
\usepackage{dsfont}
\usepackage[colorlinks=true,linkcolor=black,citecolor=black]{hyperref}
\usepackage{threeparttable}
\usepackage{array}

\usepackage{comment}
\usepackage{pdflscape}
\usepackage{afterpage}
\usepackage{subcaption}
\usepackage[labelsep=quad,skip=-2pt]{caption}

\numberwithin{equation}{section}
\theoremstyle{plain} 
\theoremstyle{plain} \newtheorem{prop}{{\sc Proposition}}
\theoremstyle{plain} 
\theoremstyle{plain} 
\theoremstyle{plain} \newtheorem{assumption}{{\sc Assumption}}

\theoremstyle{plain}

\theoremstyle{plain} 

\theoremstyle{plain}

\newcommand{\blinding}[2]{#1}   

\DeclareMathOperator\bP{\mathbb P} 
\DeclareMathOperator\bE{\mathbb E} 
\DeclareMathOperator\bV{\mathbb V} 

\newcommand{\sumi}{\sum_{i=1}^N}

\newcommand{\be}{\begin{eqnarray}}
\newcommand{\ee}{\end{eqnarray}}
\newcommand{\bee}{\begin{eqnarray*}}
\newcommand{\eee}{\end{eqnarray*}}
\newcommand{\bi}{\begin{enumerate}[(i)]}
\newcommand{\ei}{\end{enumerate}}

\begin{document}

\begin{center}
\vspace*{-2.5cm}

{\Large Generalizing trial evidence to target populations in non-nested designs: Applications to AIDS clinical trials}

\medskip
\blinding{
Fan Li$^{1,2,*}$, Ashley L. Buchanan$^3$ and Stephen R. Cole$^4$
}{}

$^1$Department of Biostatistics, Yale University School of Public Health, New Haven, CT USA\\
$^2$ Center for Methods in Implementation and Prevention Science, Yale University, New Haven, CT USA\\
$^3$Department of Pharmacy Practice, College of Pharmacy, University of Rhode Island, Kingston, RI USA\\
$^4$Department of Epidemiology, Gillings School of Public Health, University of North Carolina at Chapel Hill, Chapel Hill, NC USA\\
$*$fan.f.li@yale.edu\\

\end{center}

\date{}

{\centerline{Abstract}
\noindent Comparative effectiveness evidence from randomized trials may not be directly generalizable to a target population of substantive interest when, as in most cases, trial participants are not randomly sampled from the target population. Motivated by the need to generalize evidence from two trials conducted in the AIDS Clinical Trials Group (ACTG), we consider weighting, regression and doubly robust estimators to estimate the causal effects of HIV interventions in a specified population of people living with HIV in the USA. We focus on a non-nested trial design and discuss strategies for both point and variance estimation of the target population average treatment effect. Specifically in the generalizability context, we demonstrate both analytically and empirically that estimating the known propensity score in trials does not increase the variance for each of the weighting, regression and doubly robust estimators. We apply these methods to generalize the average treatment effects from two ACTG trials to specified target populations and operationalize key practical considerations. Finally, we report on a simulation study that investigates the finite-sample operating characteristics of the generalizability estimators and their sandwich variance estimators.
\vspace*{0.3cm}

}

\clearpage

\section{Introduction}\label{sec:intro}
\subsection{The AIDS Clinical Trials Group Studies}\label{sec:actg}
The AIDS Clinical Trials Group (ACTG) is the largest research network conducting randomized trials to study the safety and efficacy of interventions for individuals infected with human immunodeficiency virus (HIV) and those who develop acquired immunodeficiency syndrome (AIDS) \citep{Green1990}. Our motivating applications include two ACTG trials---ACTG 320 and ACTG A5202---that evaluated the efficacy of antiretroviral therapy among individuals infected with HIV. ACTG 320 compared a three-drug treatment regimen where a protease inhibitor (PI) was added to zidovudine and lamivudine with a treatment regimen only including the two nucleoside analogues, and found that adding PI significantly slowed disease progression \citep{hammer}. ACTG A5202 assessed the equivalence of abacavir-lamivudine (ABC-3TC) and tenofovir disoproxil fumarate-emtricitabine (TDF-FTC), and found that treatment with TDF-FTC was associated with lower risk of virologic failure among patients with baseline viral load $>100,000$ copies/ml \citep{sax2009abacavir, sax2011abacavir}. Because randomization balances both measured and unmeasured baseline confounders in expectation, the sample average treatment effect (SATE) represents a valid comparative parameter among the trial population \citep{Greenland1990}. However, generalizability of trial results to a broader target population living with HIV without careful considerations can be questionable because the trial population often has a different distribution of effect modifiers compared to the target population \citep{gandhi}.

We investigate whether the results of ACTG 320 and ACTG A5202 are generalizable to two target populations: all people living with HIV in the USA, and all women living with HIV in the USA. We estimate the population average treatment effect (PATE) in terms of the change in CD4 cell counts from baseline, in the absence of information on such outcomes from these target populations. Specifically, we combine the data on baseline covariates, treatments, and the outcome from the trials with baseline covariates from two cohort studies: the Center for AIDS Research Network of Integrated Clinical Systems (CNICS) \citep{Kitahata2008} and the Women's Interagency HIV Study (WIHS) \citep{Adimora2018}. For our purposes, we assume members of CNICS and WIHS are representative of all people living with HIV in the USA and all women living with HIV in the USA, respectively \citep{bacon}. Compared to the CNICS and WIHS cohorts, African-American and Hispanic women, as well as patients over 40 years were relatively under-represented in both ACTG trials \citep{buchanan}. Because race and age may be strong effect modifiers \citep{Greenbaum2008, Ribaudo2013}, the average treatment effects in trial population may be different from those in the target populations. Therefore, generalizing comparative results from the trials to the target populations requires adjusting for the differential distributions of effect modifiers between the trials and the target populations \citep{Kern2016}.

\subsection{Related Literature on Generalizability}
{Typical approaches for generalizing trial results include subclassification \citep{tipton,muirch}, outcome regression \citep{wang2019bayesian}, and inverse probability weighting \citep{cole10,stuart,hartman2015sample,buchanan}. These approaches often characterize the sampling mechanism through which patients are selected into the trials or the cohorts.} Following \cite{buchanan}, we call the probability of participation in the trial conditional on a set of covariates explaining the sampling mechanism as the \emph{sampling score}. Parallel to the treatment \emph{propensity score} which plays a central role in observational studies \citep{rosenbaum}, the sampling score represents a key quantity conditional on which the trial and population become exchangeable. In practice, the sampling score is often unknown but can be estimated. When the sampling score model is correctly specified, \citet{buchanan} demonstrated that inverse probability of sampling weighting (IPSW) outperformed subclassification, and provided a consistent sandwich variance of the IPSW estimator. In our work, we will also investigate weighting-based estimators, but do not further consider the subclassification estimators.

The IPSW generalization approach has been used in \citet{stuart} to extend experimental findings in nested trial designs, where the trial is assumed to be nested within the target population and the baseline characteristics are fully observed in the target population. Our application concerns a non-nested design where the trial sample and the sample from the target population are obtained separately \citep{dahabreh2019extending}. In particular, we assume that the characteristics of the target population are only ascertained within a random sample from the target population (the CNICS and WIHS cohorts). \citet{buchanan} proposed the IPSW approach to generalize trial results to the target population from such a non-nested design. The consistency of the IPSW estimator, however, depends on correct specification of the sampling score model. Further, it is also known that inverse weighting by the sampling score alone may not be statistically efficient \citep{Robins1994}.

To improve IPSW, \citet{Dahabreh2018a,Dahabreh2018c} considered doubly robust (DR) estimators for generalizability. Compared to IPSW, DR estimator additionally exploits the smoothness of outcome models, and provides consistent estimates to PATE if either the sampling score or the outcome model is correctly specified, but not necessarily both \citep{bang}. Frequently, we do not know which model is correct; thus, the DR estimator provides some degree of protection against model misspecification by granting two chances for valid inference. Of note, \citet{Dahabreh2018a} mainly focused on a nested trial design, but the estimators derived for the nested design may not necessarily lead to consistent estimates under the non-nested design. \citet{li2021note} and \citet{lee2021improving} have both developed calibration weighting estimators for generalizability in the nested and non-nested designs. The calibration weighting approach exploits the covariate-balancing constraints to achieve exact finite-sample balance and double robustness asymptotically. \citet{rudolph2017robust} developed targeted maximum likelihood estimators (TMLE) for transporting the intention-to-treat average effect, the average effect due to the actual treatment, and the complier average treatment effect under a non-nested design with patient noncompliance. Depending on the causal estimand, their estimator is doubly or multiply robust and may require an additional model for the actual treatment received.

{We provide a brief synthesis of the generalizability literature in Table \ref{tb:summary} to further elucidate the contribution of our article for non-nested designs. Our setting differs from \citet{cole10} and \citet{hartman2015sample} in that we observe the covariates of a large target population only through a smaller random sample. Our setting closely resembles that in \citet{buchanan}. While \citet{buchanan} considered the IPSW estimator that only depends on the sampling score, we expand their work to doubly robust estimation and explain the important role of the treatment propensity score for the purpose of generalizability. Specifically, we demonstrate both analytically and empirically that estimating the treatment propensity score can lead to notable efficiency gain for generalizing the SATE. Further, different from \citet{rudolph2017robust} and \citet{lee2021improving}, we focus on parametric modeling of the nuisance parameters, and contribute a set of computationally convenient sandwich variance estimators that account for the uncertainty in estimating the parametric models (with R code provided in Web Appendix 5). The closed-form sandwich variances for the DR estimators, for example, have not been explicitly derived in the generalizability literature. Finally, we also present a comprehensive case study to assess the generalizability of two ACTG trials and operationalize key considerations for estimating PATE and its corresponding sampling variance.}

\begin{center}
[Table \ref{tb:summary} about here.]
\end{center}

\section{Notation and Assumptions}\label{sec:setup}

Suppose the scientific interest lies in drawing inference about the effect of a time-fixed binary treatment on an outcome measured at the end of follow-up in a target population. We assume that each individual in the target population has a pair of potential outcomes $\{Y^0, Y^1\}$ \citep{Rubin1978}. Here $Y^0$ is the outcome that would have been observed if, possibly contrary to fact, the individual received ``usual care'' (e.g., the conventional treatment regimen with two nucleoside analogues in ACTG 320), and $Y^1$ is the outcome that would have been seen if the individual received ``treatment'' (e.g., the treatment regimen that also included the PI). Define $\mu_1 = \bE(Y^1)$ and $\mu_0 =\bE(Y^0)$ as the average potential outcomes in the target population. The causal parameter of interest is the PATE, defined as
\begin{equation*}
\Delta =\bE(Y^1-Y^0)=\mu_1-\mu_0.
\end{equation*}
To estimate $\Delta$, we consider a scenario with two sources of available information: a sample of $n$ individuals from the target population who participate in a trial (i.e., one of the ACTG trials), and a cohort of $m$ individuals (i.e., the CNICS or WIHS study) randomly drawn from the target population. While the cohort is assumed to be representative of the target population, the trial participants often differ from the non-randomized individuals in important ways. We further assume the knowledge of the size of the actual population from which the cohort participants are sampled, which is sufficient for estimating parameters in the sampling score model. We do not require that the trial participants are nested in the cohort sample; this is a practical consideration because individual identifiers are frequently unavailable.

Throughout we make the Stable Unit Treatment Value Assumption (SUTVA), which implies treatment variation irrelevance and no interference. The assumption of treatment variation irrelevance holds if the same version of treatment could be provided to all trial participants and (potentially) non-participants in the target population, or if differences among versions of treatment (such as delivery mechanism) are irrelevant to the outcome of interest \citep{vanderweele2009further}. This assumption may be invalid, for example, when the treatment administration in the trial is accompanied by adherence counseling, while such counseling is absent when treatment is provided to the non-randomized participants. The absence of interference means that the potential outcome of each individual does not depend on the treatment received by others \citep{hudgens2008toward}. This assumption may be questionable, for example, in a vaccine trial, where the vaccination status of one individual may affect whether another individual develops flu due to herd immunity.

Let $Z$ be a $p$-vector of baseline covariates observed for both the trial and cohort participants. Define $S=1$ if the individual participates in the trial and $S=0$ otherwise. For trial participants, define $X$ as the treatment indicator, with $X=1$ indicating active treatment and $X=0$ otherwise. Under SUTVA, the observed outcome for each trial participant is $Y=Y^1 X+Y^0(1-X)$, while we do not observe either treatment or outcome within the cohort. We also define $D$ as an indicator for inclusion in the study, where $D=1$ implies the individual is included in the observed data (combined trial or cohort sample) and $D=0$ otherwise. Furthermore, we assume that if $S=1$, then $D=1$. In short, we observe information on $(Z, S=1,X,Y,D=1)$ for trial participants, and information on $(Z,S=0,D=1)$ for the cohort sample. {Notice that our set up differs from \citet{wang2019bayesian} in that we assume neither $Y$ nor $X$ are observed in the cohort study (CNICS or WIHS). In cases when both $Y$ and $X$ are available in multiple cohort studies, we refer to \citet{wang2019bayesian} who developed a Bayesian nonparametric outcome regression estimator for integrative analysis of trials and cohorts.} In our setting, as we do not observe all potential outcomes in the target population, the identification and inference for $\Delta$ requires the following two assumptions.

\begin{assumption}\label{asp:random}(Randomization) The treatment is randomly assigned in the trial, namely
\be\label{eq:random}
\bP(X=1|S=1,Z, Y^0, Y^1)=\bP(X=1|S=1).
\ee
The randomization probability $\bP(X=1|S=1)= r \in (0,1)$.
\end{assumption}

\begin{assumption}\label{asp:SI}(Ignorable Trial Participation) Conditional on the set of covariates $Z$, trial participation is independent of the potential outcomes, namely,
\be\label{eq:SI}
\bP(S=1|Z, Y^0, Y^1)=\bP(S=1|Z).
\ee
The sampling score, defined as $w(Z)=\bP(S=1|Z)$, is strictly positive for all $Z$ with a positive density.
\end{assumption}

Typically, the treatment is randomly assigned in the trial and Assumption \ref{asp:random} holds by design. In this case, the true treatment propensity score is $e(W)=\bP(X=1|S=1,W)=r$, for any subset of baseline covariates $W\subseteq Z$. Assumption \ref{asp:SI} requires exchangeability between trial participants and non-participants conditional on the pre-treatment covariates $Z$. Assumption \ref{asp:SI} further requires positivity in trial participation such that there is a positive probability of participating in the trial for each value of the covariates \citep{westreich2010invited}. Although positivity can be checked by visualizing the distribution of the estimated sampling scores \citep{stuart}, the conditional exchangeability \eqref{eq:SI} is not testable and merits sensitivity analysis \citep{Nguyen2017}. Finally, we assume the absence of noncompliance such that the treatment actually received by each individual is the same as the randomized treatment within the trial. In the presence of treatment noncompliance, there exist alternative causal estimands to describe the within-trial treatment effect, including the per-protocol causal effect and the complier average causal effect. Additional structural assumptions and different statistical strategies are required to enable identification of these alternative within-trial causal estimands, prior to generalizations to new target populations. We refer to \citet{rudolph2017robust} and \citet{lu2019generalizing} for more explicit definitions of alternative estimands and generalization strategies in the presence of noncompliance.

\section{Estimating Population Average Treatment Effect}
\label{sec:estimation}
\subsection{Preliminaries}\label{sec:prelim}
We consider five estimators of $\Delta$ to generalize trial results to a specified target population. We assume a finite-dimensional logistic model for the sampling scores $w(Z;\gamma)=\bP(S=1|Z;\gamma) = \lbrace 1+\exp(-Z^T \gamma)\rbrace^{-1}$, where $\gamma$ is a $p$-vector of regression coefficients. Throughout we assume the vector $Z$ includes 1 as the first component to accommodate an intercept. Let $\widehat{\gamma}$ denote the weighted maximum likelihood estimator of $\gamma$ where each trial participant is given weight $\Pi_{1}^{-1}=1$ and each member in the cohort is given weight $\Pi_{0}^{-1}=(N-n)/m$, and $N$ is the target population size \citep{scott1986fitting}. Because the weight depends only on trial participation status, we write $\Pi_{S_i}^{-1}=S_i\Pi_{1}^{-1}+(1-S_i)\Pi_{0}^{-1}$. In our setting, the trial sample is much smaller than the target population, and we could reasonably approximate the inclusion probability $\pi_0=\bP(D=1|S=0)\approx m/(N-n)$, which motivates the choice of weight $\Pi_0^{-1}$. {The intuition behind $\Pi_0^{-1}$ is that it \emph{replaces} each cohort participant by $(N-m)/m$ copies of him or herself to fully represent the $N-m$ non-randomized participants in the target population, so that all $N$ observations in the target population are used to estimate $w(Z)$ without conditioning on $D=1$.} On the other hand, if the cohort sample coincides with the target population, then $m = N$ and $\Pi^{-1}_0= (m-n)/m \approx 1$ and our estimators approximate those developed for the nested trial design \citep{stuart,Dahabreh2018a}. For each observation in the trial and the cohort, we denote the estimated sampling score by $\widehat{w}_i= w({Z_i};\widehat{\gamma})$.

Although the true treatment propensity score is known in the trials, estimating the propensity score (rather than the using the true value) may lead to improved efficiency of the SATE estimator, by controlling for chance imbalance \citep{Rosenbaum1987,Hirano2003}. For a set of baseline covariates $W\subseteq Z$, we use a \emph{working} logistic model for the propensity scores $e(W;\beta)=\bP(X=1|S=1,W;\beta) = \lbrace 1+\exp(-W^T\beta)\rbrace^{-1}$, where $\beta$ is a $q$-vector of coefficients. Specifically, one could include prognostic covariates or covariates that exhibit baseline imbalance into $W$. In this case, we write $\widehat{\beta}$ as the maximum likelihood estimator, and the estimated propensity scores $\widehat{e}_{i}=e(W_i;\widehat{\beta})$. \citet{Dahabreh2018a} recommended using the estimated propensity scores in nested trial designs, but did not provide an mathematical justification. To strengthen that recommendation, we establish asymptotic results in Section \ref{sec:asym3} and show that the potential efficiency gain due to estimating propensity scores applies to all five generalizability estimators we consider below.

\subsection{Inverse Probability of Sampling Weighting}\label{sec:IPSW}

We first consider two IPSW estimators; both estimators weight the trial sample by the inverse of the estimated sampling score to approximate the covariate distribution in the target population. If the sampling score model is correctly specified, both estimators remove the bias due to non-random trial participation and provide consistent estimates of PATE. With the estimated sampling scores and treatment propensity scores, the first IPSW estimator is akin to a Horvitz-Thompson estimator in survey sampling, and is written as
\begin{equation}
\label{deltahat1}
\widehat{\Delta}_{\text{IPSW1}} =  \frac{1}{N}\sumi \frac{D_iS_iX_iY_i}{\widehat{w}_i\widehat{e}_{i}}
-\frac{1}{N}\sumi \frac{D_iS_i(1-X_i)Y_i}{\widehat{w}_i(1-\widehat{e}_{i})}.
\end{equation}
The second IPSW estimator is akin to the H\'{a}jek estimator, and is given by
\begin{equation}
\label{deltahat2}
\widehat{\Delta}_{\text{IPSW2}} =  \frac{\sumi D_iS_iX_iY_i/\widehat{w}_i\widehat{e}_{i}}{\sumi D_iS_iX_i/\widehat{w}_i\widehat{e}_{i}}
-\frac{\sumi D_iS_i(1-X_i)Y_i/\widehat{w}_i(1-\widehat{e}_i)}{\sumi D_iS_i(1-X_i)/\widehat{w}_i(1-\widehat{e}_i)}.
\end{equation}
In the causal inference literature, $\widehat{\Delta}_{\text{IPSW2}}$ typically produces an effect estimate within the range of the observed outcomes and may be more efficient than $\widehat{\Delta}_{\text{IPSW1}}$. {Because $\bP(D=1|S=1)=1$, the inclusion indicator $D_i$ can be omitted from $\widehat{\Delta}_{\text{IPSW1}}$ and $\widehat{\Delta}_{\text{IPSW2}}$. We include $D_i$ in our presentation throughout because, as will be seen in Section \ref{sec:asym}, this notation allows us to treat the collection of all random variables as independent and identically distributed (IID) copies from the target population of size $N$, allowing us to invoke the standard asymptotic theory for IID data.} Both $\widehat{\Delta}_{\text{IPSW1}}$ and $\widehat{\Delta}_{\text{IPSW2}}$ use an estimated treatment propensity score $\widehat{e}_i$. When the true treatment propensity score is used, however, the propensity score factors out of the IPSW2 estimator, and we obtain the estimator studied by \citet{buchanan}. We denote the corresponding estimators obtained by replacing $\widehat{e}_i$ with the true propensity score as $\widetilde{\Delta}_{\text{IPSW1}}$ and $\widetilde{\Delta}_{\text{IPSW2}}$.

\subsection{Outcome Regression}\label{sec:REG}
For the outcome regression estimator, we write $m_x(Z)=\bE(Y^x|Z,S=1)$ for the conditional expectation of the potential outcome among trial participants under intervention $x$. By SUTVA, $m_x(Z)=\bE(Y|Z,X=x,S=1)$, and we could posit parametric models for these conditional expectations $m_x(Z;\alpha_x)$, $x=0,1$, where $\alpha_1$ and $\alpha_0$ are $l_1\times 1$ and $l_0\times 1$ vectors of regression coefficients, respectively. In general, we could obtain maximum likelihood estimators, $\widetilde{\alpha}_1$ and $\widetilde{\alpha}_0$, as solutions to the following estimating equations
\begin{align}
&\sumi D_iS_iX_i\psi_{\alpha_1}(Y_i,Z_i;\alpha_1)=0\label{eq:om1}\\
&\sumi D_iS_i(1-X_i)\psi_{\alpha_0}(Y_i,Z_i;\alpha_0)=0\label{eq:om0}
\end{align}
where $\psi_{\alpha_1}(Y_i,Z_i;\alpha_1)$ and $\psi_{\alpha_0}(Y_i,Z_i;\alpha_0)$ are score functions determined by model specification. For example, if we use a linear model $m_1(Z)=Z^T\alpha_1$, then $\psi_{\alpha_1}(Y_i,Z_i;\alpha_1)=Z_i(Y_i-Z_i^T\alpha_1)$. {Notice that equations \eqref{eq:om1} and \eqref{eq:om0} correspond to a strategy that fits a separate regression model within each treatment group. In the presence of treatment-by-covariate interactions, this strategy obviates the need to estimate treatment-by-covariate interactions, but is identical to fitting a single regression model in the trial sample including full treatment-by-covariate interactions \citep{lunceford,Dahabreh2018c}.} We write the predicted outcome for each individual as $\widetilde{m}_{1i}=m_1(Z_i;\widetilde{\alpha}_1)$, $\widetilde{m}_{0i}=m_0(Z_i;\widetilde{\alpha}_0)$, and define the outcome regression (REG) estimator of the PATE as
\begin{equation}
\label{deltatildeo1}
\widetilde{\Delta}_{\text{REG}} = \frac{1}{N}\sumi c_i(\widetilde{m}_{1i}-\widetilde{m}_{0i}).
\end{equation}
{where $c_i=D_i\lbrace S_i + \Pi_0^{-1}(1-S_i) \rbrace$ is a population standardization factor that depends on the inclusion probability. Intuitively, because the cohort sample only includes $m$ observations, the factor $\Pi_0^{-1}$ {replaces} each cohort participant by $(N-m)/m$ copies of him or herself to fully represent the $N-m$ non-randomized participants in the target population.} By this intuition, we could also see that $\widetilde{\Delta}_{\text{REG}}$ is consistent for $\Delta$ when the outcome models are correctly specified.

Alternatively, we consider an outcome analysis assisted by the estimated propensity scores. To do that, we obtain $\widehat{\alpha}_1$ and $\widehat{\alpha}_0$ from the observed data by solving the following weighted estimating equations
\begin{align}
&\sumi D_iS_iX_i\psi_{\alpha_1}(Y_i,Z_i;\alpha_1)/\widehat{e}_i=0,\label{eq:wom1}\\
&\sumi D_iS_i(1-X_i)\psi_{\alpha_0}(Y_i,Z_i;\alpha_0)/(1-\widehat{e}_i)=0.\label{eq:wom0}
\end{align}
Because the true propensity score is known by design, $\widehat{e}_i$ is always correctly specified and estimating equations \eqref{eq:wom1} and \eqref{eq:wom0} are unbiased as long as \eqref{eq:om1} and \eqref{eq:om0} are unbiased. Now write the predicted outcome for each individual as $\widehat{m}_{1i}=m_1(Z_i;\widehat{\alpha}_1)$, $\widehat{m}_{0i}=m_0(Z_i;\widehat{\alpha}_0)$, the REG estimator becomes \begin{equation}
\label{deltahato1}
\widehat{\Delta}_{\text{REG}} = \frac{1}{N}\sumi c_i(\widehat{m}_{1i}-\widehat{m}_{0i}).
\end{equation}
In particular, the above REG estimator differs from the standard REG estimator \eqref{deltatildeo1} as we have introduced inverse probability of treatment weighting. In fact, $\widehat{e}_i$ is not required for the consistency of the REG estimator, and we can treat $\widetilde{\Delta}_{\text{REG}}$ as the version of \eqref{deltahato1} where the estimated propensity score is replaced by the truth $e_i=r$. In this case, as the randomization probability is constant, the propensity score term factors out of the estimating equations \eqref{eq:wom1} and \eqref{eq:wom0}. From this perspective, $\widehat{e}_i$ seems redundant for performing outcome regression. However, as we explain in Section \ref{sec:asym3}, the use of $\widehat{e}_i$ does not increase the asymptotic variance; in other words, $\widehat{\Delta}_{\text{REG}}$ is at least as efficient as $\widetilde{\Delta}_{\text{REG}}$.

\subsection{Doubly Robust Estimators}
We additionally consider two DR estimators that combine IPSW and regression. Based on Assumptions \ref{asp:random} and \ref{asp:SI}, we show in Web Appendix 1 that the efficient influence function for estimating the PATE {in our non-nested design} is
\begin{align}
\mathcal{I}_{\text{eff}}(D_i,S_i,X_i,Y_i,Z_i)=&\frac{D_iS_iX_i}{w_ie_i}(Y_i-m_1(Z_i))-
\frac{D_iS_i(1-X_i)}{w_i(1-e_i)}(Y_i-m_0(Z_i))\nonumber\\
&+D_i\{S_i+\pi_0^{-1}(1-S_i)\}(m_1(Z_i)-m_0(Z_i)-\Delta),
\end{align}
where $w_i$, $e_i$ are the true sampling score and the propensity score, {and the population standardization factor, $D_i\{S_i+\pi_0^{-1}(1-S_i)\}$, depends on the true inclusion probability $\pi_0$}. Replacing $w_i$, $e_i$ with estimated $\widehat{w}_i$, $\widehat{e}_i$, and $m_1(Z_i)$, $m_0(Z_i)$ with the estimated $\widehat{m}_{1i}$, $\widehat{m}_{0i}$, the solution of $\Delta$ based on $\sumi \mathcal{I}_{\text{eff}}(D_i,S_i,X_i,Y_i,Z_i)=0$ motivates the first DR estimator
\begin{align}
\label{deltahatDR1}
\widehat{\Delta}_{\text{DR1}}=&\frac{1}{N} \sumi  \bigg\{ \frac{D_iS_i X_i}{\widehat{w}_i\widehat{e}_{i}}(Y_i-\widehat{m}_{1i})
-\frac{D_iS_i(1-X_i)}{\widehat{w}_i(1-\widehat{e}_{i})}(Y_i-\widehat{m}_{0i})
+c_i(\widehat{m}_{1i}-\widehat{m}_{0i})
  \bigg\},
\end{align}
{where $c_i=D_i\lbrace S_i + \Pi_0^{-1}(1-S_i) \rbrace$ is a population standardization factor that depends on the estimated inclusion probability $\Pi_0$}. By construction, $\widehat{\Delta}_{\text{DR1}}$ is an IPSW1 estimator augmented by outcome regression. Similar to $\widehat{\Delta}_{\text{IPSW1}}$, the inverse probability of sampling weights in $\widehat{\Delta}_{\text{DR1}}$ is unbounded and therefore motivates the application of H\'{a}jek weights to construct the second DR estimator,
\begin{align}
\label{deltahatDR2}
\widehat{\Delta}_{\text{DR2}} =& \frac{\sumi D_iS_iX_i(Y_i-\widehat{m}_{1i})/\widehat{w}_i\widehat{e}_{i}}{\sumi D_iS_iX_i/\widehat{w}_i\widehat{e}_{i}}
- \frac{\sumi D_iS_i(1-X_i)(Y_i-\widehat{m}_{0i})/\widehat{w}_i(1-\widehat{e}_{i})}{\sumi D_iS_i(1-X_i)/\widehat{w}_i(1-\widehat{e}_{i})}\nonumber\\
&+ \frac{1}{N} \sumi c_i(\widehat{m}_{1i}-\widehat{m}_{0i}).
\end{align}
The estimators $\widehat{\Delta}_{\text{DR1}}$ and $\widehat{\Delta}_{\text{DR2}}$ use the estimated propensity score $\widehat{e}_i$, but we can again replace that with the known propensity score $e_i=r$ and define the corresponding estimators $\widetilde{\Delta}_{\text{DR1}}$ and $\widetilde{\Delta}_{\text{DR2}}$. Because the true propensity score is a constant, this term also factors out of $\widetilde{\Delta}_{\text{DR2}}$.

In Web Appendix 2, we confirm that, in our non-nested design, both $\widehat{\Delta}_{\text{DR1}}$ and $\widehat{\Delta}_{\text{DR2}}$ converge to $\Delta$ in large samples, as long as either the sampling score model or the outcome models are correctly specified but not necessarily both. This property provides two opportunities for valid generalizability analyses, and renders the DR estimators potentially more attractive over IPSW and REG alone. When all models are correctly specified, the DR estimators are also more efficient than IPSW alone \citep{Robins1994}. Of note, in a nested trial design where $m=N$, we have $\Pi_{0}^{-1}\approx 1$ for the cohort sample and therefore $\widehat{\Delta}_{\text{DR1}}$ and $\widehat{\Delta}_{\text{DR2}}$ reduce to the DR estimators in \citet{Dahabreh2018a}. Finally, the robustness property of $\widehat{\Delta}_{\text{DR1}}$ and $\widehat{\Delta}_{\text{DR2}}$ suggests a practical approach to diagnose model misspecification \citep{robins2001comment,MercatantiLi2014}. That is, assuming visual assessment of the sampling score distribution suggests no violation of positivity in trial participation, if DR estimate is different from the REG estimate, but is close to IPSW estimate, it suggests potentially misspecified outcome models. On the other hand, if the DR estimate is close to REG estimate but is different from IPSW, then the model for the probability of trial participation may be misspecified.

\section{Large-Sample Properties and Efficiency Considerations}\label{sec:asym}

We express each generalizability estimator as the solution to unbiased estimating equations to establish the consistency and asymptotic normality in the non-nested design setting. We focus on DR2, and considerations for other estimators are given in Web Appendix 3. {The unbiased estimating equations representation permits the derivation of a sandwich variance estimator that accounts for the the uncertainty in estimating the parametric nuisance models. Of note, these additional sources of uncertainty were considered important for accurate variance estimation when parametric models are used to estimate the nuisance parameters in observational studies \citep{lunceford,buchanan,LiandLi2019,mao2019propensity}, and we extend such considerations to these generalizability estimators.} In the following, we assume the random vectors, $(D_i,D_iS_i,D_iZ_i,D_iS_iX_i,D_iS_iY_i)$, $i=1,\ldots,N$ are IID draws from the target population. Our asymptotic analysis requires the target population size $N$ to approach infinity, and as $N\rightarrow \infty$, the inclusion probability approaches a positive constant: $\Pi_0=m/(N-n)\rightarrow \pi_0$. Similar assumptions are used in the choice-based sampling literature \citep{scott1986fitting,li2020secondary}

\subsection{Large-Sample Distribution of $\widetilde{\Delta}_{\text{DR2}}$}\label{sec:asymp1}
We first consider the case where only the sampling score and the outcome models are estimated, and the true propensity score is used. As the sampling scores are estimated by weighted maximum likelihood, $\widehat{\gamma}$ solves the $p \times 1$ estimating equation
\begin{align*}
\sumi \psi_{\gamma}(D_i,S_i,{Z}_i;\gamma) =\sumi
\frac{D_i\Pi_{S_i}^{-1}(S_i-w_i)}{w_i( 1-w_i)}\frac{\partial}{\partial {\gamma} }w_i = {0},
\end{align*}
\noindent {which is obtained from differentiating the weighted binomial log-likelihood of the sampling score model with respect to $\gamma$.} The estimation of parameters, $\widetilde{\alpha}_1$ and $\widetilde{\alpha}_0$ in the outcome model are based on solving equations \eqref{eq:om1} and \eqref{eq:om0}, with the use of the true propensity scores. We denote the probability limits of these parameter estimates ($\widehat{\gamma}$, $\widetilde{\alpha}_1$ and $\widetilde{\alpha}_0$) as $\gamma^*$, $\alpha_1^*$ and $\alpha_0^*$. We use the star superscripts because the respective models may be misspecified, and thus $\gamma^*$, $\alpha_1^*$ and $\alpha_0^*$ are probability limits of the potentially misspecified models, which are allowed to be different from the parameter values in their respective true models \citep{White1982}.

{In Web Appendix 3.5, we write $\widetilde{\Delta}_{\text{DR2}}=\widetilde{\nu}_1-\widetilde{\nu}_2+\widetilde{\nu}_3$, where $\widetilde{\nu}_1$, $\widetilde{\nu}_2$, $\widetilde{\nu}_3$ correspond to the three summands in \eqref{deltahatDR2} but with $\hat{e}_i$ omitted and $\widehat{m}_{1i}$, $\widehat{m}_{0i}$ replaced by $\widetilde{m}_{1i}$, $\widetilde{m}_{0i}$. Let $\widetilde{\theta}=(\widetilde{\nu}_1,\widetilde{\nu}_2,\widetilde{\nu}_3,\widehat{\gamma}^T,\widetilde{\alpha}_1^T,\widetilde{\alpha}_0^T)^T$, and define $\theta^*=(\nu_1,\nu_2,\nu_3,\gamma^{*T},\alpha_1^{*T},\alpha_0^{*T})^T$ as the limiting value of $\widetilde{\theta}$.} Then $\widetilde{\theta}$ is the solution for $\theta^*$ in the $(3+p+l_1+l_0)\times 1$ estimating equation $\sumi \Psi_{\Delta_{\text{DR2}}}(Y_i,D_i,S_i,X_i,Z_i;\widetilde{\theta})=0$, where
\begin{align}\label{eq:psi}
\small
&\Psi_{\Delta_{\text{DR2}}}(Y_i,D_i,S_i,X_i,Z_i;\theta) =
\begin{pmatrix}
D_iS_iX_i(Y_i-m_{1i}-\nu_1)/(w_ie_i)\\
D_iS_i(1-X_i)(Y_i-m_{0i}-\nu_2)/(w_i(1-e_i))\\
c_im_{1i}-c_im_{0i}-\nu_3\\
\psi_{\gamma}(S_i,Z_i;\gamma)\\
D_iS_iX_i\psi_{\alpha_1}(Y_i,Z_i;\alpha_1)/e_i\\
D_iS_i(1-X_i)\psi_{\alpha_0}(Y_i,Z_i;\alpha_0)/(1-e_i)
\end{pmatrix}.
\end{align}
Define $A(\theta^*)=\bE\left\{\frac{\partial}{\partial\theta^T}
{\Psi}_{\Delta_{\text{DR2}}}(Y_i,D_i,S_i,X_i,Z_i;\theta^*)\right\}$ and ${B}(\theta^*)= \bV\left\{{\Psi}_{\Delta_{\text{DR2}}}(Y_i,D_i,S_i,X_i,Z_i;\theta^*)\right\}$,
where the expectation and covariance operators are defined with respect to the target population. The fact that the joint estimating equations are unbiased, i.e., $\bE[{\Psi}_{\Delta_{\text{DR2}}}(Y_i,D_i,S_i,X_i,Z_i;\theta^*)]=0$, indicates that under suitable regularity conditions, as $N\rightarrow \infty$, $N^{1/2}(\widetilde{\theta}-\theta^*)$ converges in distribution to $N(0,\Omega_{\theta^*})$, where ${\Omega}_{\theta^*}=A(\theta^*)^{-1}B(\theta^*)A(\theta^*)^{-T}$ \citep{Stefanski2002}. By an application of Slutsky's theorem and the delta method, $\widetilde{\Delta}_{\text{DR2}}$ is a consistent estimator of $\Delta$ and $N^{1/2}(\widetilde{\Delta}_{\text{DR2}}-\Delta)$ converges in distribution to $N(0,{\sigma}^2_{\text{DR2}})$ where ${\sigma}^2_{\text{DR2}}=\lambda^T{\Omega}_{\theta^*}\lambda$ and $\lambda=(1,-1,1,0_{1\times p},0_{1\times l_1},0_{1\times l_0})^T$, $0_{r\times c}$ is a $r\times c$ matrix of zeros. A consistent sandwich variance estimator for $\widetilde{\Delta}_{\text{DR2}}$ is then given by
\begin{equation}
\widehat{\bV}(\widetilde{\Delta}_{\text{DR2}})=\frac{1}{N}\lambda^T A(\widetilde{\theta})^{-1}B(\widetilde{\theta})A(\widetilde{\theta})^{-T}\lambda.
\end{equation}

\subsection{Large-Sample Distribution of $\widehat{\Delta}_{\text{DR2}}$}
With an estimated treatment propensity score $\widehat{e}_i=e(W_i;\hat{\beta})$, we can write the estimator $\widehat{\beta}$ as the solution to the $q \times 1$ estimating equation
\begin{align*}
\sumi \psi_{\beta}(D_i,S_i,{X}_i,{W}_i;\beta) =\sumi
\frac{D_iS_i(X_i-e_i)}{e_i(1-e_i)}\frac{\partial e_i}{\partial\beta} = {0},
\end{align*}
\noindent {which is obtained from differentiating the binomial log-likelihood of the treatment propensity score model with respect to $\beta$. In Web Appendix 3.5, we write $\widehat{\Delta}_{\text{DR2}}=\widehat{\nu}_1-\widehat{\nu}_2+\widehat{\nu}_3$, where $\widehat{\nu}_1$, $\widehat{\nu}_2$, $\widehat{\nu}_3$ correspond to the three summands in \eqref{deltahatDR2}, respectively. Let $\widehat{\varpi}=(\widehat{\nu}_1,\widehat{\nu}_2,\widehat{\nu}_3,\widehat{\gamma}^T,\widehat{\alpha}^T_1,\widehat{\alpha}^T_0,\widehat{\beta}^T)^T$, and define $\varpi^{*}=(\theta^{*T},\beta^T)^T$ as the limiting value of $\widehat{\varpi}$.} Then $\widehat{\varpi}$ is the solution to the $(3+p+l_1+l_0+q)\times 1$ estimating equation $\sumi \Phi_{\Delta_{\text{DR2}}}(Y_i,D_i,S_i,X_i,Z_i;\widehat{\varpi}) =0$, where
\begin{align*}
\small
& \Phi_{\Delta_{\text{DR2}}}(Y_i,D_i,S_i,X_i,Z_i;\varpi) =
\begin{pmatrix}
{\Psi}_{\Delta_{\text{DR2}}}(Y_i,D_i,S_i,X_i,Z_i;\varpi)\\
\psi_{\beta}(D_i,S_i,{X}_i,{W}_i;\beta)
\end{pmatrix},
\end{align*}
where $\Psi_{\Delta_{\text{DR2}}}(Y_i,D_i,S_i,X_i,Z_i;\varpi)=\Psi_{\Delta_{\text{DR2}}}(Y_i,D_i,S_i,X_i,Z_i;\theta)$ whenever $\beta$ is chosen such that $e(W_i,\beta)=e_i$, the true propensity score. Define ${C}(\varpi^{*})=\bE\left\{\frac{\partial}{\partial\varpi^T}
{\Phi}_{\Delta_{\text{DR2}}}(Y_i,D_i,S_i,X_i,Z_i;\varpi^{*})\right\}$,
${D}(\varpi^{*})= \bV\left\{{\Phi}_{\Delta_{\text{DR2}}}(Y_i,D_i,S_i,X_i,Z_i;\varpi^{*})\right\}$. Notice that $\bE[{\Phi}_{\Delta_{\text{DR2}}}(Y_i,D_i,S_i,X_i,Z_i;\varpi^{*})]=0$, which indicates that under suitable regularity conditions, as $N\rightarrow\infty$, $N^{1/2}(\widehat{\varpi}-\varpi^{*})$ converges in distribution to $N(0,{\Omega}_{\varpi^{*}})$, where ${\Omega}_{\varpi^{*}}={C}(\varpi^{*})^{-1}{D}(\varpi^{*})
{C}(\varpi^{*})^{-T}.$ Further, we must have $\widehat{\Delta}_{\text{DR2}}$ is a consistent estimator of $\Delta$ and $N^{1/2}(\widehat{\Delta}_{\text{DR2}}-\Delta)$ converges in distribution to $N(0,{\tau}^2_{\text{DR2}})$ where ${\tau}^2_{\text{DR2}}=\eta^T{\Omega}_{\varpi^{*}}{\eta}$ and ${\eta}=(\lambda^T,0_{1\times q})^T$. A consistent sandwich variance estimator for $\Delta_{\text{DR2}}$ is therefore
\begin{equation}
\widehat{\bV}(\widehat{\Delta}_{\text{DR2}})=\frac{1}{N}\eta^T C(\widehat{\varpi})^{-1}D(\widehat{\varpi})C(\widehat{\varpi})^{-T}\eta.
\end{equation}

\subsection{Analytical Comparison}\label{sec:asym3}
An analytical comparison between ${\sigma}^2_{\text{DR2}}$ and ${\tau}^2_{\text{DR2}}$ reveals that the asymptotic variance is guaranteed to be no larger when the treatment propensity score is estimated, because
\begin{equation}\label{eq:ineq}
{\tau}^2_{\text{DR2}}={\sigma}^2_{\text{DR2}}-\{\lambda^TA(\theta^*)G^T\} E^{-1}_{\beta\beta}\{\lambda^TA(\theta^*)G^T\}^T\leq {\sigma}^2_{\text{DR2}},
\end{equation}
where $G$ is the lower left off-diagonal block of $B(\theta^*)$ (defined in Web Appendix 3) and $E_{\beta\beta}=\bE[D_iS_i\{\partial e_i/\partial\beta\}\{\partial e_i/\partial\beta\}^T/(e_i(1-e_i))]$ is a positive definite Hessian matrix. {The efficiency gain due to estimating known propensity scores is a classic result for the inverse probability weighting estimator in observational studies \citep{robins1992estimating,Hirano2003,wooldridge2007inverse}, and \eqref{eq:ineq} extends the same result to doubly robust generalizability estimators. We can also repeat the above derivation for the rest of the four generalizability estimators, and show that inequality \eqref{eq:ineq} still holds true, without requiring the sampling score or outcomes models to be correctly specified. Proposition \ref{prop:eff} below summarizes this comparative finding, with technical details given in Web Appendix 3. Although we show these inequality results by directly deriving and comparing the asymptotic variances, we remark that an alternative proof can proceed by treating $e_i$ as the sole nuisance parameter and verifying the tangent space condition in Theorem 3 of \citet{hitomi2008puzzling} for each estimator.}

\begin{prop}\label{prop:eff}
Suppose the propensity score is modeled by a smooth parametric model $e(W;\beta)$, and the parameters are estimated by maximum likelihood. The asymptotic variances of $\widehat{\Delta}_{\text{IPSW1}}$, $\widehat{\Delta}_{\text{IPSW2}}$, $\widehat{\Delta}_{\text{REG}}$, $\widehat{\Delta}_{\text{DR1}}$, $\widehat{\Delta}_{\text{DR2}}$ do not exceed those of $\widetilde{\Delta}_{\text{IPSW1}}$, $\widetilde{\Delta}_{\text{IPSW2}}$, $\widetilde{\Delta}_{\text{REG}}$, $\widetilde{\Delta}_{\text{DR1}}$, $\widetilde{\Delta}_{\text{DR2}}$, respectively.
\end{prop}

Interestingly, there exists a stronger version of Proposition \ref{prop:eff} stating that including additional baseline covariates in the propensity score model will not compromise the asymptotic efficiency of the generalizability estimator. This result is summarized in Proposition \ref{prop:eff2}, and the proof is provided in Web Appendix 4.

\begin{prop}\label{prop:eff2}
Suppose $W_1$ and $W_2$ are two sets of baseline covariates, and $W_1\subseteq W_2$, and let $e(W_1;\beta_1)$ and $e(W_2;\beta_2)$ be smooth nested parametric models in the sense that there exists $\xi(\beta_1)$ such that $e(W_1;\beta_1)=e(W_2;\beta_1,\xi(\beta_1))$ for every $\beta_1$, $W_1$, $W_2$. If $\widehat{\beta}_1$, $\widehat{\beta}_2$ are estimated by maximum likelihood, and $e(W_1;\widehat{\beta}_1)$, $e(W_2;\widehat{\beta}_2)$ are the corresponding estimated propensity scores, then the five generalizability estimators constructed with $e(W_2;\widehat{\beta}_2)$ are asymptotically at least as efficient as their counterparts constructed with  $e(W_1;\widehat{\beta}_1)$.
\end{prop}

Proposition \ref{prop:eff2} grants the use of a more saturated treatment propensity score model with extra covariates because this strategy does not reduce the large-sample variance of the generalizability estimators. The insight is that estimating a more saturated propensity scores serves as an implicit step to adjust for more baseline covariates in trials. Although it has been shown that adjusting additional prognostic covariates improves the efficiency of the SATE estimator \citep{Moore2011,shen2014inverse,williamson2014variance,Zeng2020}, the role of such adjustment for purpose of generalizability has not been fully articulated analytically in the generalizability literature such as those in Table \ref{tb:summary}. Proposition \ref{prop:eff2} bridges this gap by concluding that improved efficiency in estimating the SATE through implicit covariate adjustment, can translate into potentially improved efficiency in estimating the PATE, across all five generalizability estimators.

\section{Assessing Generalizability of ACTG 320 and ACTG A5202}\label{sec:app}
\subsection{Trial and Cohort Data}

The ACTG 320 trial enrolled participants between January 1996 and January 1997 and examined the efficacy of adding a PI to an HIV treatment regimen with two nucleoside analogues \citep{hammer}. Around 20\% of the 1,156 participants were women. We focus on the change in the CD4 cell counts as the outcome of interest, and a treatment regimen is favored if it leads to an increase in CD4 cell count from baseline (i.e., reflects improvements in immunological functioning). At week 4 follow-up, 116 (10\%) patients had missing CD4 cell count and were excluded from this analysis. The baseline characteristics of the excluded patients were not systematically different from the remaining patients, and therefore we only considered a complete-case analysis to focus on the generalizability aspect of the problem. The ACTG A5202 trial enrolled participants between September 2005 and November 2007 and assessed the equivalence of ABC-3TC or TDF-FTC plus efavirenz or ritonavir-boosted atazanavir \citep{sax2009abacavir,sax2011abacavir}. About 17\% of the 1,857 participants were women. At week 48 follow-up, 417 (22\%) patients had missing CD4 cell count and were likewise excluded. Web Tables 1 to 4 summarize the baseline characteristics of all participants and the women subgroup by study arms. In both trials, the majority of participants are between 30 to 40 years old. While the majority (53\% and 41\%) of all participants in ACTG 320 and ACTG 5202 were non-Hispanic White, the majority (46\% and 55\%) of all women participants in ACTG 320 and ACTG 5202 were African-American.

For this analysis, we assume the CNICS and WIHS cohorts to be representative samples from their respective target populations, namely, all people living with HIV in the USA, and all women living with HIV in the USA. With comprehensive clinical data from point-of-care electronic medical record systems for population-based HIV research, the CNICS cohort includes over 27,000 HIV-infected adults from eight USA CFAR sites \citep{Kitahata2008}. On the other hand, the WIHS cohort includes 4,129 women recruited from six USA sites, and represents the oldest prospective cohort study of women with and at risk for HIV infection \citep{bacon,Adimora2018}. We respectively harmonize the two cohorts so that the final cohort samples each match the key eligibility criteria in the two ACTG trials. In particular, we selected the first record for a participant in the cohort study that met key study inclusion criteria specific to each trial. For generalizing ACTG 320, we restrict the cohort sample to those who were HIV-positive, highly active antiretroviral therapy (HAART) naive, and had CD4 cell counts lower than 200 cells/mm$^3$ at the previous visit. This leads to $m=6,158$ participants from CNICS and $m=493$ women from WHIS. For generalizing ACTG A5202, we restrict the cohort sample to those who were HIV-positive, antiretroviral therapy (ART) naive, and had viral load greater than 1000 copies/ml at the previous visit. This provides $m=12,302$ participants from CNICS and $m=1,012$ women from WIHS. Baseline characteristics of the resulting cohort samples, including calendar time of visit, time since ART initiation, age, CD4 cell count, and viral load, are summarized in Web Tables 5 and 6. Finally, based on the \citet{cdcest} estimates, the size of the first target population is assumed to be $N=1.1$ million (all people living with HIV), and the size of the second target population is assumed to $N=280,000$ (women living with HIV).

\subsection{Model Specifications and Balance Check}
For each analysis, the combined ACTG trial and cohort sample is used to fit a weighted logistic model and estimate the sampling scores. The sampling score model includes variables associated with selection into the trial or treatment effect modifiers with a linear term for continuous variables, as well as all pairwise interactions. Sex, race, age, history of injection drug use (IDU), and baseline CD4 are included in the sampling score model for generalizing ACTG 320, while sex, race, age, history of IDU, hepatitis B or C, AIDS diagnosis, baseline CD4 and baseline viral load (on log$_{10}$ scale) are included in the sampling score model for generalizing ACTG A5202. We also incorporated a squared age term and its interactions with other covariates because this strategy improves weighted covariate balance. Sex is excluded in the sampling score model when generalizing the trial results among the women subgroup.

\begin{figure}[htbp]
  \begin{subfigure}{1\textwidth}
    \centering
    \caption{\emph{Trial (320) and cohort (CNICS)}}
    \includegraphics[scale=0.65]{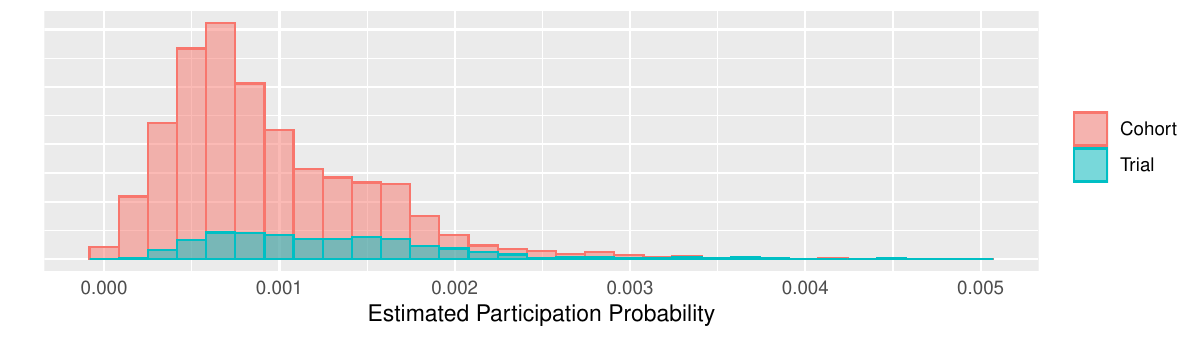}
  \end{subfigure}
  \begin{subfigure}{1\textwidth}
    \centering
    \caption{\emph{Trial (A5202) and cohort (CNICS)}}
    \includegraphics[scale=0.65]{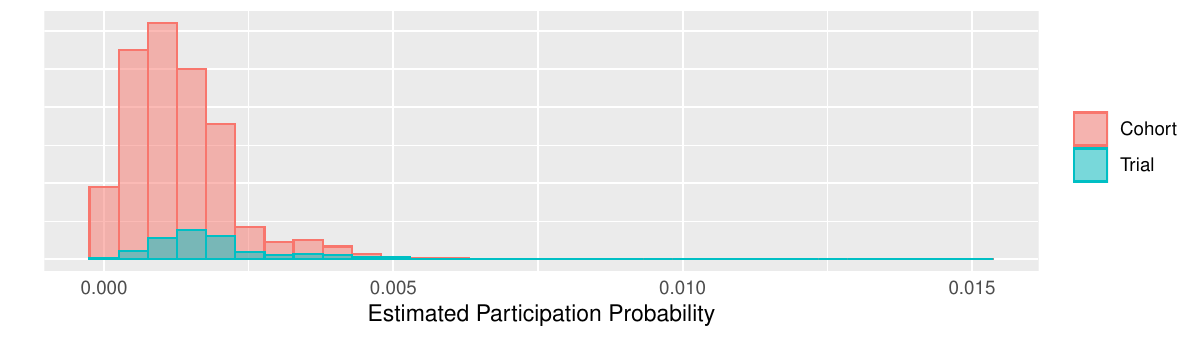}
  \end{subfigure}
  \begin{subfigure}{1\textwidth}
    \centering
    \caption{\emph{Trial (320) and cohort (WIHS)}}
    \includegraphics[scale=0.65]{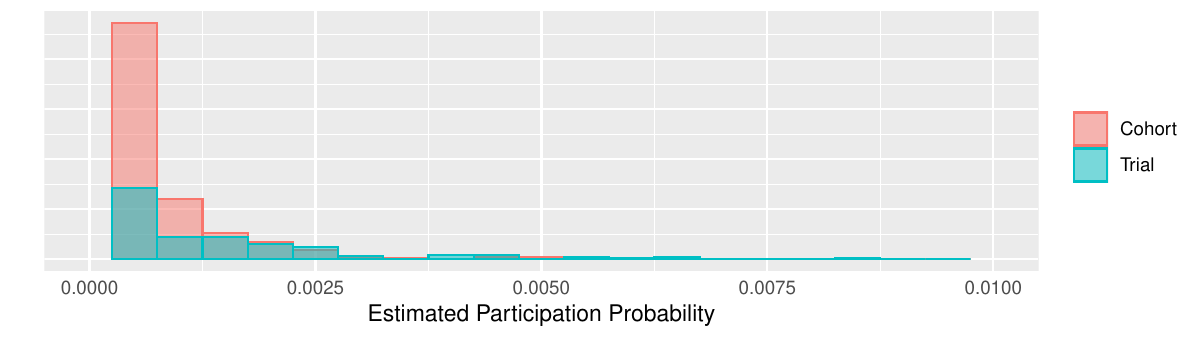}
  \end{subfigure}
  \begin{subfigure}{1\textwidth}
    \centering
    \caption{\emph{Trial (A5202) and cohort (WIHS)}}
    \includegraphics[scale=0.65]{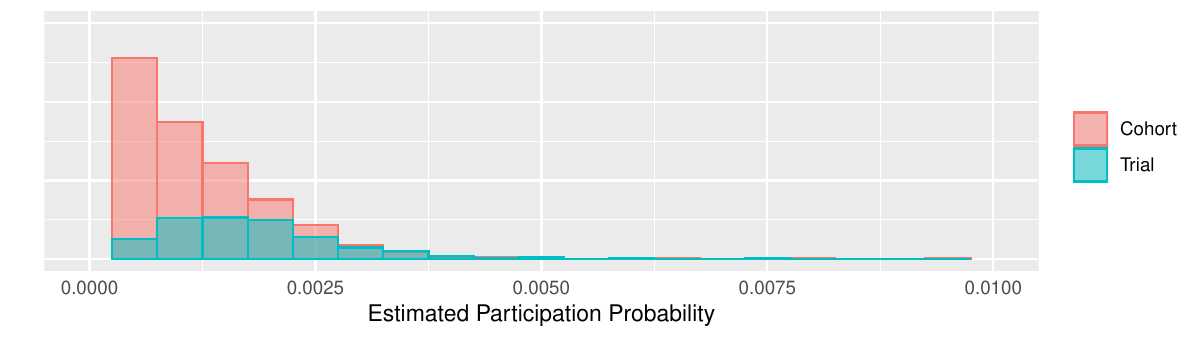}
  \end{subfigure}
  \caption{\emph{Histograms of estimated sampling scores for each of the four generalizability analyses.}}\label{fig:hist0}
\end{figure}

\begin{figure}[ht]
  \begin{subfigure}{.5\textwidth}
    \centering
    \caption{\emph{Trial (320) and cohort (CNICS)}}
    \includegraphics[scale=0.65]{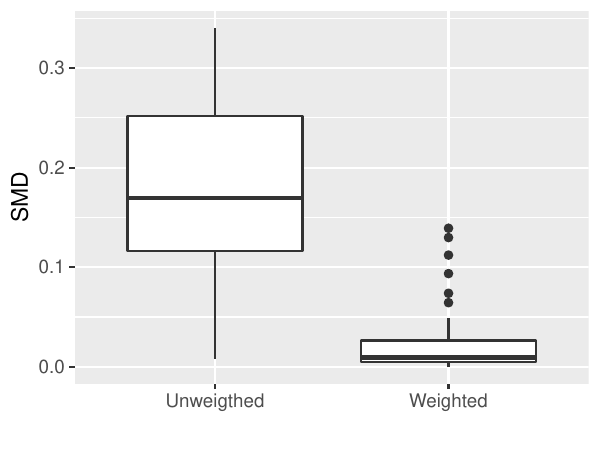}
  \end{subfigure}
  \begin{subfigure}{.5\textwidth}
    \centering
    \caption{\emph{Trial (A5202) and cohort (CNICS)}}
    \includegraphics[scale=0.65]{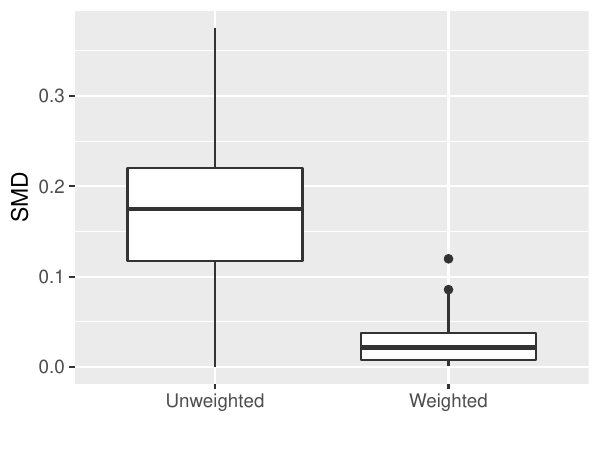}
  \end{subfigure}
  \begin{subfigure}{.5\textwidth}
   \centering
   \caption{\emph{Trial (320) and cohort (WIHS)}}
   \includegraphics[scale=0.65]{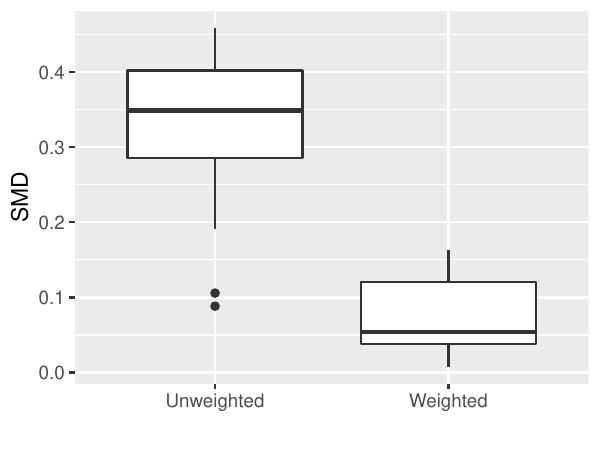}
  \end{subfigure}
  \begin{subfigure}{.5\textwidth}
    \caption{\emph{Trial (A5202) and cohort (WIHS)}}
    \centering\includegraphics[scale=0.65]{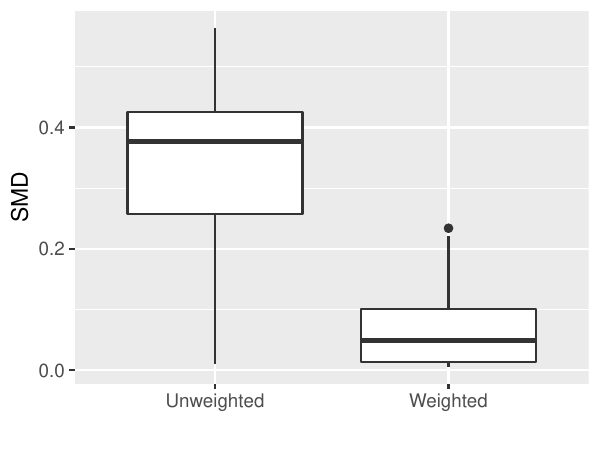}
  \end{subfigure}
  \caption{\emph{Boxplots of the standard mean differences (SMDs) of all covariates (and interaction terms) for the unweighted trial sample and inverse probability of participation weighted trial sample for each of the four analyses.}}\label{fig:box0}
\end{figure}

Figure \ref{fig:hist0} presents the histograms of estimated sampling scores. The histograms facilitate a visual check on the positivity assumption for trial participation. Even though the magnitude of the sampling scores is generally small due to the large population size $N$, the distribution of sampling scores between the trial and cohort do not signal a strong lack of common support (also see Web Figure 1 for a zoomed version of the tails for each histogram). To further check the adequacy of the estimated sampling scores, we calculate the balance for each covariate between the weighted sample and population. {Extending the definition of \citet{Austin2015}, we define the standardized mean difference (SMD) for non-nested design as}
\begin{equation*}
\text{SMD}=\frac{1}{s}\left\vert\frac{\sumi D_iS_iZ_{i(k)}\pi_i}{\sumi D_iS_i\pi_i}-\frac{\sumi D_i\lbrace S_i + \Pi_0^{-1}(1-S_i) \rbrace Z_{i(k)}}{N}\right \vert,
\end{equation*}
where $Z_{i(k)}$ is the $k$th regressor included in the sampling score model. The denominator $s$ is the standard deviation of $Z_{i(k)}$ in the target population and estimated by
\begin{equation*}
s=\left\{\frac{mN\sumi D_i\lbrace S_i + \Pi_0^{-1}(1-S_i) \rbrace (Z_{i(k)}-\bar{Z}_{(k)})^2}{m(N^2-n)-(N-n)^2}
\right\}^{1/2},
\end{equation*}
where $\bar{Z}_{(k)}={\sumi D_i\lbrace S_i + \Pi_0^{-1}(1-S_i) \rbrace Z_{i(k)}}/{N}$ is the population average. This expression is essentially the weighted standard deviation of each covariate with weights $\Pi^{-1}_{S_i}$. When $\pi_i=1$, $\text{SMD}$ quantifies the systematic difference between the trial and population, and reflects the degree of trial sample selection bias. When $\pi_i=\widehat{w}_i^{-1}$, the $\text{SMD}$ measures the similarity between the weighted trial and population, and is used as a diagnostic check of the sampling score weights. Figure \ref{fig:box0} summarizes the SMD of all covariates before and after sampling score weighting across the four analyses. In Web Figures 2 to 5, we also provide the forest plot of the SMD by each covariate for each generalizability analysis. It is evident that the sampling score weights improve balance between trial and population, with the largest $\text{SMD}$ controlled under $20\%$ after weighting (an exception is when generalizing A5202 to WHIS, where the SMD for baseline CD4 and one interaction term with baseline CD4 are slightly above 20\%). While a more stringent balance threshold (10\%) has been previously suggested for analyzing observational studies \citep{Austin2015}, here we use 20\% as a less stringent threshold due to the large pre-existing differences in the trial and population sizes.

\begin{center}
[Figure \ref{fig:hist0} about here.]
\end{center}

\begin{center}
[Figure \ref{fig:box0} about here.]
\end{center}

We retain the same set of covariates in the outcome model as in the sampling score model. Linear regression including main effects and all pairwise interactions were fit among the trial participants. We follow the strategy in Section \ref{sec:REG} and fit separate outcome models within each treatment group, before predicting the unobserved potential outcomes for the entire observed sample. In particular, among women in ACTG A5202, there are no participants in the control group with both hepatitis B/C and an AIDS diagnosis, so this interaction term is excluded. The quantile-quantile plots of regression residuals do not suggest violations of the normality assumption and are omitted for brevity.

We consider both the true propensity scores ($e_i=0.5$) and the estimated propensity scores ($\widehat{e}_i$) in constructing the estimators for PATE. To estimate the treatment propensity scores, we specify a logistic regression of treatment on a set of baseline covariates $W_i$. Two strategies are used to specify $W_i$. The first strategy only includes the main effects of baseline covariates and represents a more parsimonious specification (referred to as \emph{main-effects logistic model}). Information of baseline covariates by treatment group in the trials are summarized in Web Tables 1 to 4. The second strategy includes the main effects and pairwise interactions of all baseline covariates in $W_i$ (referred to as \emph{full logistic model}). Due to sparse cell counts, we include race as binary variable (White vs. non-White) in the propensity score model when generalizing ACTG A5202 to all women living with HIV in the USA (WIHS cohort). In this particular analysis, the following pairwise interactions are also excluded from the full logistic propensity score model to avoid numerical issues with sparse cell counts: race-hepatitis, race-AIDS, race-IDU, squared-age-IDU, hepatitis-AIDS, and IDU-AIDS. Compared to the analyses using the true treatment propensity scores, Proposition \ref{prop:eff} indicates that the above two strategies of specifying $W_i$ can control for important baseline covariates and improve the precision of the PATE estimators. Furthermore, Proposition \ref{prop:eff2} suggests there is no asymptotic efficiency loss by over-specifying the propensity score model, as in the second strategy. For each analysis, we estimate the variance and associated 95\% confidence interval for PATE using the proposed sandwich variance estimator. The study protocol was reviewed and approved by the University of Rhode Island Institutional Review Board.

\subsection{Assessing Generalizability}\label{sec:app3}

The SATE in each trial is estimated by the difference-in-means estimator (Table \ref{tableall_k}). In ACTG 320, there is a notable improvement in the CD4 cell response from baseline to 4 weeks among patients included in the PI group compared to those in the non-PI group, both overall (SATE = 19) and for the women subgroup (SATE=24). These changes are statistically significant at the 5\% level as the 95\% confidence intervals (CIs) are $(12,25)$ and $(7,41)$, respectively. In ACTG A5202, those randomized to ABC-3TC had slightly higher average change in CD4 cell count from baseline to week 48 compared to those randomized to TDF-FTC (SATE=6). Among the women subgroup, the two treatment groups had a similar average change in CD4 (SATE=1). The results in ACTG A5202 are not statistical significant at the 5\% level as the respective 95\% CIs are $(-8,20)$ and $(-35,37)$.

\begin{center}
[Table \ref{tableall_k} about here.]
\end{center}

\begin{figure}[htbp]
  \begin{subfigure}{.5\textwidth}
    \centering
    \caption{Trial (320) and cohort (CNICS)}
    \includegraphics[scale=0.4]{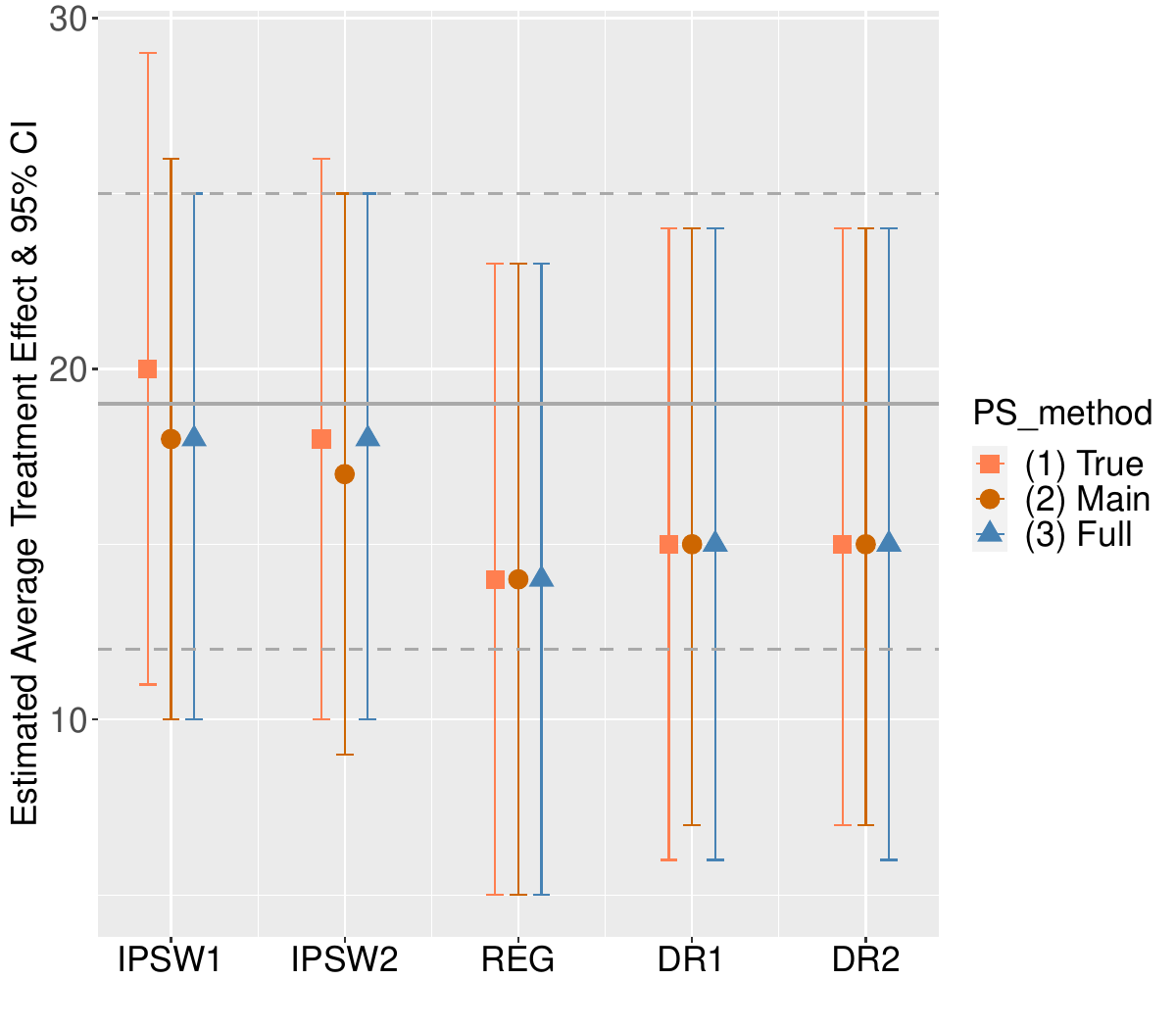}
  \end{subfigure}
  \begin{subfigure}{.5\textwidth}
    \centering
    \caption{Trial (A5202) and cohort (CNICS)}
    \includegraphics[scale=0.4]{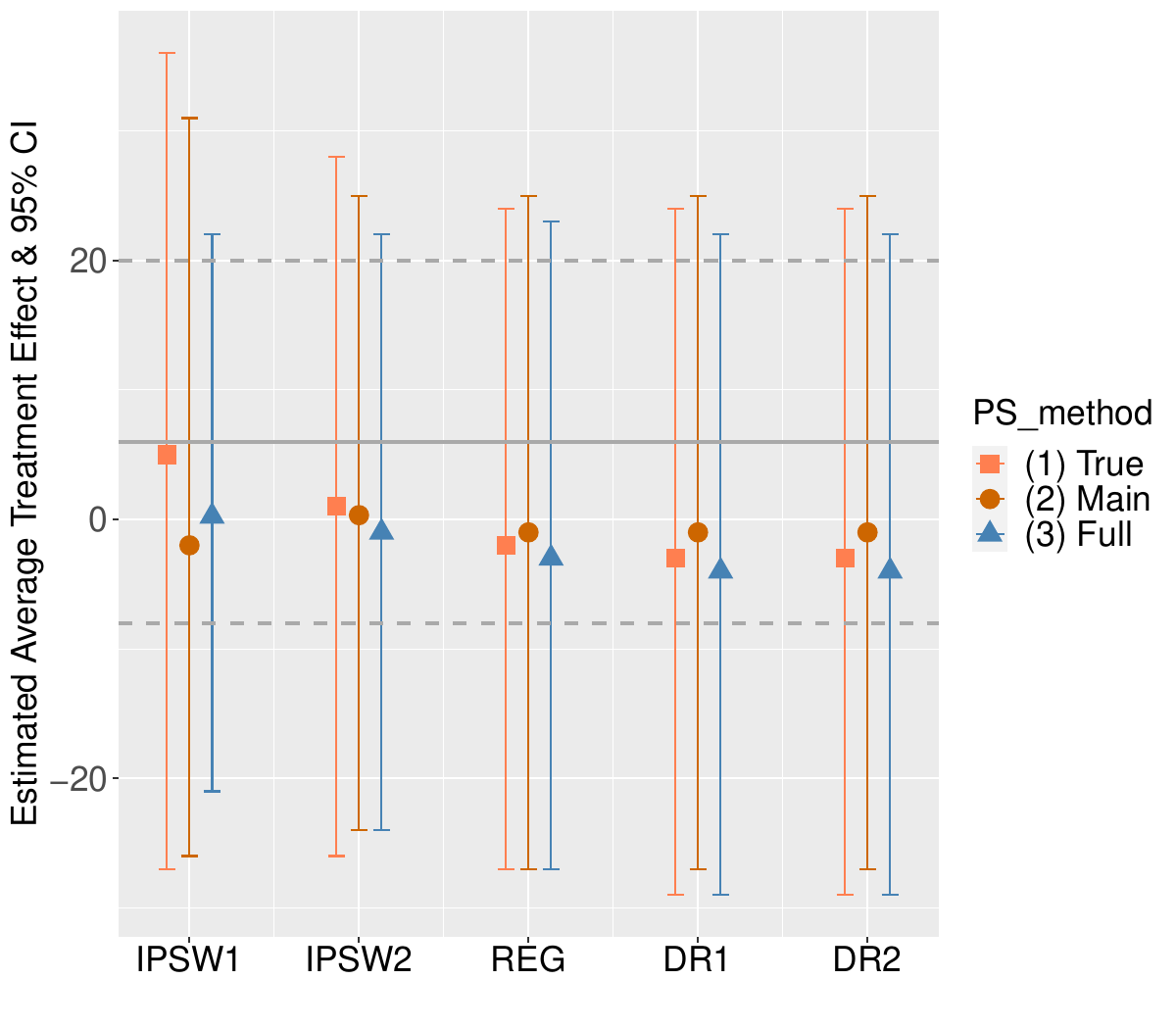}
  \end{subfigure}
  \begin{subfigure}{.5\textwidth}
   \centering
   \caption{Trial (320) and cohort (WIHS)}
   \includegraphics[scale=0.4]{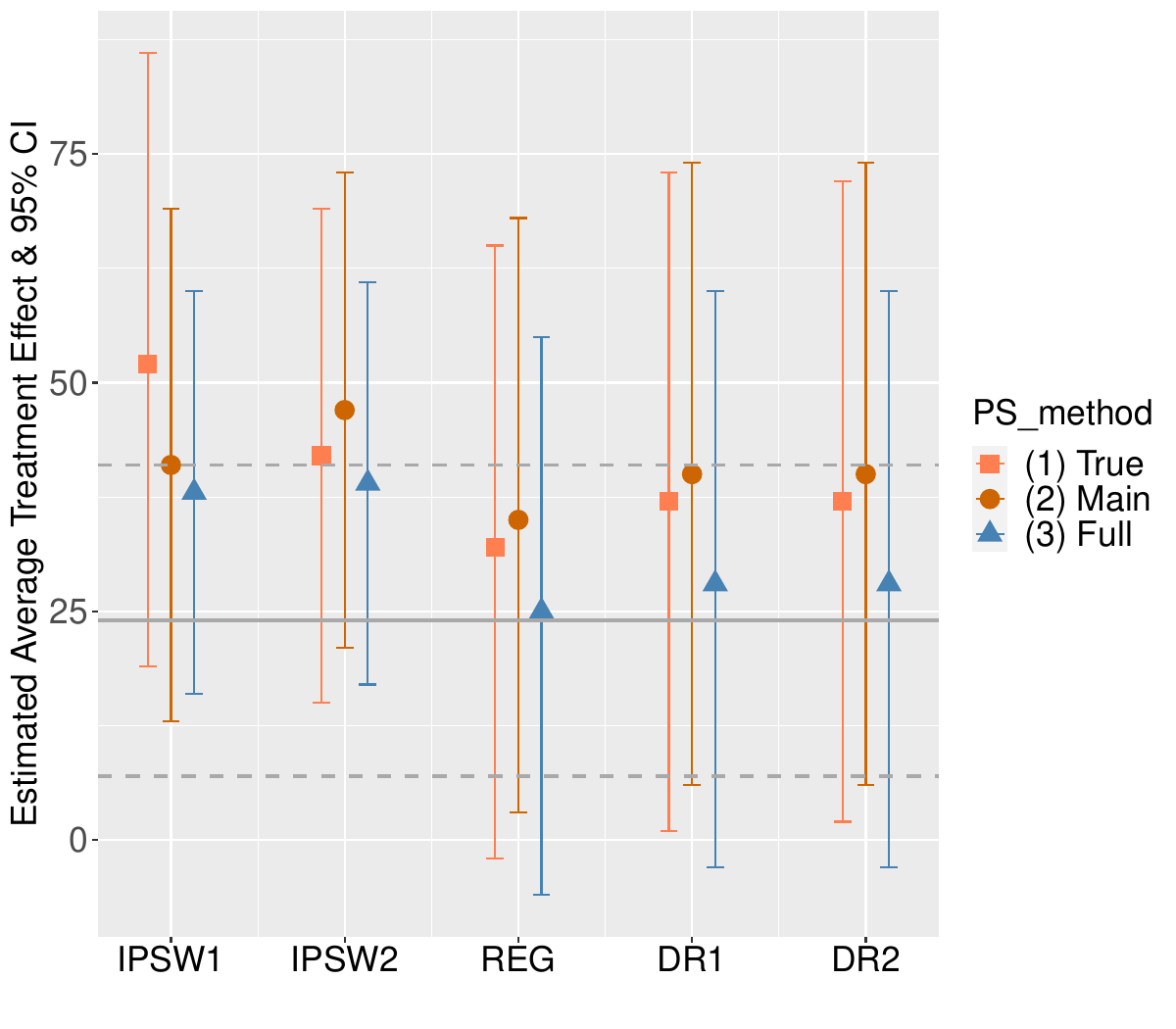}
  \end{subfigure}
  \begin{subfigure}{.5\textwidth}
    \caption{Trial (A5202) and cohort (WIHS)}
    \centering\includegraphics[scale=0.4]{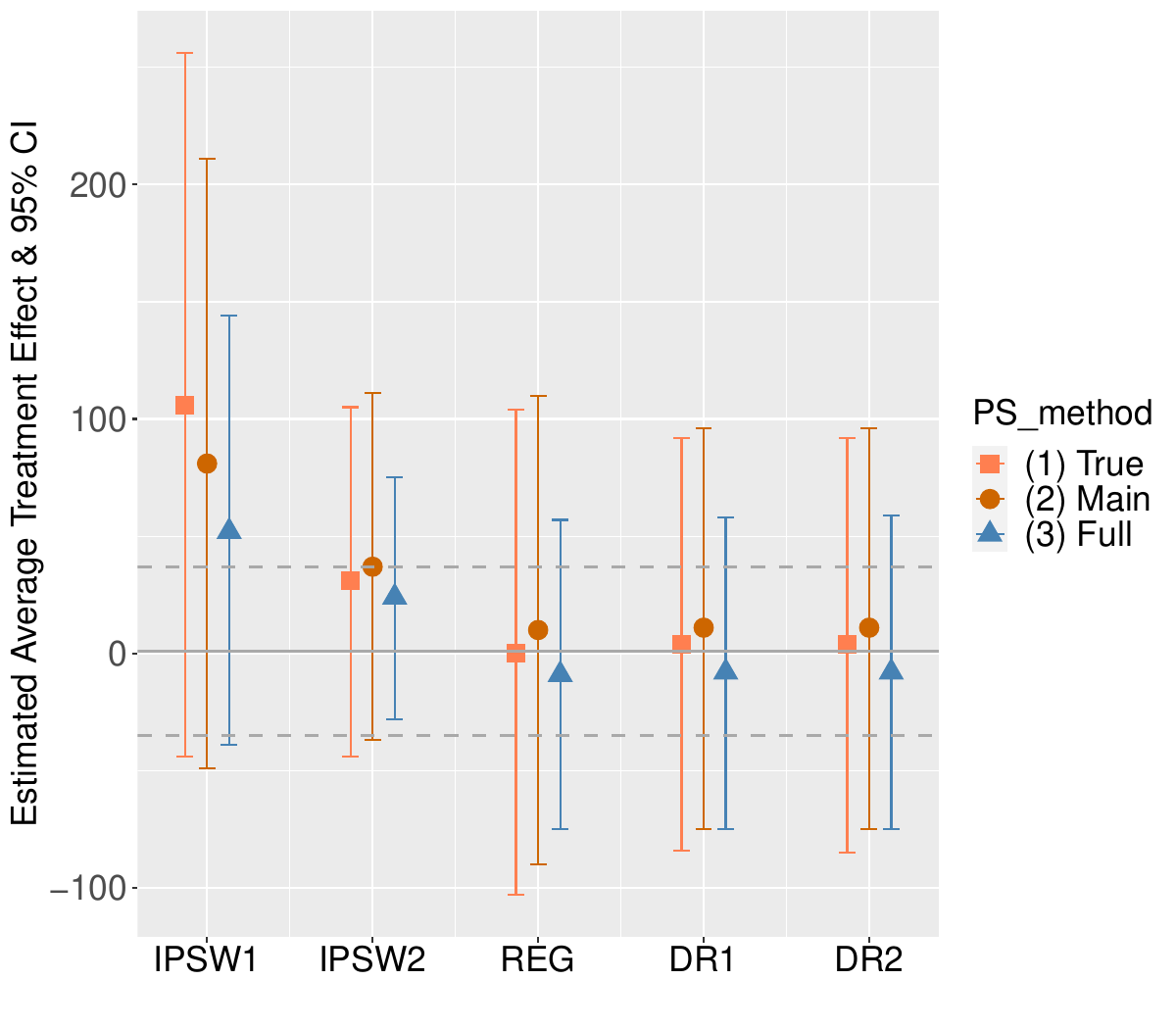}
  \end{subfigure}
  \caption{PATE estimates and 95$\%$ confidence intervals. The three sets of results correspond to generalizability estimators that use (i) the true propensity score (PS method$=$True); (ii) the estimated propensity score using a main-effects logistic model (PS method$=$Main); and (iii) the estimated propensity score using a full logistic model (PS method$=$Full). The dashed lines indicate results for the within-trial SATE.}\label{fig:analysis}
\end{figure}

Table \ref{tableall_k} summarizes the PATE estimates in both target populations generalized from each ACTG trial. The analysis is conducted using both the true propensity score and the estimated propensity scores; all CIs are based on the sandwich variance estimators developed in Section \ref{sec:asym} and Web Appendix 3. We also present the results in Figure \ref{fig:analysis}, facilitating a graphical comparison across the three sets of results obtained with different treatment propensity score estimates. First, the three sets of results in Table \ref{tableall_k} and Figure \ref{fig:analysis} empirically illustrates the asymptotic findings in Proposition \ref{prop:eff} and \ref{prop:eff2} in that the CIs obtained with the main-effects logistic propensity score model are generally no wider than those obtained with the true propensity score, and the CIs obtained with the full logistic propensity score model are often the narrowest. The differences in widths of CI can be substantial for each of the five estimators, with the largest differences observed when generalizing ACTG A5202 to all women living with HIV. In the following, we mainly interpret the results with the full logistic propensity score model as they appear to be the most efficient.

\begin{center}
[Figure \ref{fig:analysis} about here.]
\end{center}

For generalizing the ACTG 320 to the target population of all people living with HIV, the PATE estimate is generally similar to the SATE estimate, regardless of the use of IPSW, REG or DR approaches; for example, PATE is estimated as $\widehat{\Delta}_{\text{IPSW2}}=18$, $\widehat{\Delta}_{\text{REG}}=14$ and $\widehat{\Delta}_{\text{DR2}}=15$. The generalization analysis slightly inflates variance and leads to wider CIs. Specifically, the 95\% CIs of PATE are $(10,25)$, $(5,23)$ and $(6,24)$ using the IPSW2, REG and DR2 methods, suggesting that the combination with PI is likely to induce positive CD4 response in the target population with a comparable magnitude as in ACTG 320. For generalizing ACTG A5202 to the target population of all people living with HIV in the USA, while the SATE estimate is positive, the PATE estimates are mostly negative. The 95\% CIs for PATE are fairly symmetric around the null; for example, the 95\% CIs of PATE are $(-24,22)$, $(-27,23)$ and $(-29,22)$ using the IPSW2, REG and DR2 methods. Overall, the effect of PI in ACTG 320 may be more generalizable to the target population of all people living with HIV in the USA, whereas the effect the ART combination ABC-3TC (versus TDF-FTC) may be less generalizable to this same target population.

In the target population of all women living with HIV in the USA, the PATE estimates obtained by IPSW are approximately 1.6 times the SATE estimate from ACTG 320 ($\widehat{\Delta}_{\text{IPSW1}}=38$, $\widehat{\Delta}_{\text{IPSW2}}=39$ compared to SATE$=24$). The REG estimate appears closer to SATE ($\widehat{\Delta}_{\text{REG}}=25$), and the two DR estimates fall in between IPSW and REG estimates. Using the full logistic propensity score model, while the two CI estimates from IPSW exclude zero (95\% CI is $(16,60)$ from IPSW1 and $(17,61)$ from IPSW2), the CI estimates from REG and DR contain zero (95\% CI is $(-6,55)$ from REG and $(-3,60)$ from DR1 and DR2). On the other hand, all PATE CIs exclude zero when using the main-effects logistic propensity score model; for example, the 95\% CI of PATE is $(3,68)$ from REG and $(6,74)$ from DR1 and DR2. The magnitudes of the PATE estimates using either propensity score model specification suggest that the within-trial SATE estimate may slightly underestimate the treatment effect of PI for all HIV-infected women in the USA. When generalizing ACTG A5202 to the target population of all women living with HIV, the PATE estimates using IPSW suggest a much greater CD4 cell count increase from baseline compared to the SATE estimate. While IPSW1 provides the largest point estimate ($\widehat{\Delta}_{\text{IPSW1}}=52$) with a much wider CI $(-39,144)$, the PATE estimates obtained with other approaches are smaller and had narrower CIs; the 95\% CIs for PATE are $(-28,75)$ under IPSW2, $(-75,57)$ under REG, $(-75,58)$ under DR1 and $(-75,59)$ under DR2. Among them, the IPSW2 estimate still indicates a protective effect of ABC-3TC (versus TDF-FTC) in the target population ($\widehat{\Delta}_{\text{IPSW2}}=24$), but this estimated effect is attenuated and becomes negative when using REG and DR with the full logistic propensity score model. Likewise, the PATE estimates are attenuated by both REG and DR when using the main-effects logistic propensity score model, though they remain positive. In this analysis, because the DR estimates are closer to REG than to IPSW, the results signify a potentially misspecified sampling score model. However, the estimated 95\% CI for PATE from each method contains zero, and therefore the PATE is not significantly different from null at the 5\% level. The differences between the SATE and PATE estimates imply that results from ACTG A5202 may not be directly generalizable to all women living with HIV.

Finally, even though we consider the \citet{cdcest} estimates as the best guesses for the target population size $N$, we also carried out additional analyses to assess the sensitivity of the PATE estimates to different assumptions of $N$. In particular, we specify $N\in\{0.7\text{ million},1.5\text{ million}\}$ when the target population is all people living with HIV and with $N\in \{230000,330000\}$ when the target population is all women living with HIV. The former represents a scenario where the population size estimate is off by $0.4$ million and the latter where the population size estimate is off by $50000$, in either direction. The point and interval estimates for PATE are presented in Web Tables 7 and 8. For this application, we observe that the PATE estimates obtained by each method were generally insensitive to the specified larger and smaller sizes of the target populations.

\section{Simulation Studies}\label{sec:sim}
\subsection{Main Simulation Design}\label{sec:sim1}
We conduct a simulation study that mimics the motivating setting to further elucidate the comparison between IPSW, REG and DR estimators in scenarios with two effect modifiers and a continuous outcome. We generate one binary covariate $Z_{i1}\sim\text{Bernoulli}(0.4)$ and one continuous covariate $Z_{i2}$ from $N(0,1)$. We assume a target population of size $N=1$ million, where the true sampling score is $w_i=\lbrace 1 + \exp(-\gamma_0-\gamma_1 Z_{1i}-\gamma_2 Z_{i2}-\gamma_3 Z_{i1}Z_{i2})\rbrace^{-1}$. A Bernoulli trial participation indicator, $S_i$, is generated based on $w_i$ and only those with $S_i=1$ participate in the trial. Among the trial participants, the treatment indicator is randomized with $X_i\sim \text{Bernoulli}(0.5)$; the potential outcomes are generated from
\begin{align*}
Y_i^1=& \alpha_{10} + \alpha_{11} Z_{1i} +\alpha_{12} Z_{i2}+\alpha_{13} Z_{i1}Z_{i2}+\epsilon_{1i},\\
Y_i^0=& \alpha_{00} + \alpha_{01} Z_{1i} +\alpha_{02} Z_{i2}+\alpha_{03} Z_{i1}Z_{i2}+\epsilon_{0i},
\end{align*}
where $\epsilon_{1i}$, $\epsilon_{0i}$ are independent $N(0,1)$ error terms. We choose $\alpha_{10}=2$, $\alpha_{01}=\alpha_{02}=\alpha_{03}=-1$ and $\alpha_{00}=0$, and vary the values of $\alpha_{11}$, $\alpha_{12}$, $\alpha_{13}$ to represent different levels of effect modification. Define $\zeta_1=\alpha_{11}-\alpha_{01}$, $\zeta_2=\alpha_{12}-\alpha_{02}$, $\zeta_3=\alpha_{13}-\alpha_{03}$ and $Z_{i1}$, $Z_{i2}$, $Z_{i1}Z_{i2}$ will be considered as effect modifiers as long as the association parameters $\zeta=(\zeta_1,\zeta_2,\zeta_3)^T$ are nonzero. The sampling score model parameters $\gamma=(\gamma_0,\gamma_1,\gamma_2,\gamma_3)^T$ are chosen such that the trial size is $n\approx 1,000$. We also simulate a cohort as a random sample of size $m=4,000$ from the target population (less those selected into the trial). Similar to the motivating application, the number of participants in the trial is small compared to the size of the target, and the cohort can be considered as a simple random sample from the target population. Furthermore, this choice of sample sizes resembles those for generalizing the two ACTG trials to the population represented by the CNICS cohort.

We consider four scenarios with two sets of different values of $\gamma$ and two sets of values of $\zeta$. We choose $\gamma=(-7.148, 0.3,0.3,0.3)^T$ and $\gamma=(-7.698, 0.6,0.6,0.6)^T$ to represent moderate and strong selection effect in trial participation, and set $\zeta=(1,1,1)^T$, $\zeta=(2,2,2)^T$ to denote moderate and strong effect modification. We calculate the true PATE for each scenario based on the distribution of $Z_{i1}$, $Z_{i2}$ in the target population, and obtain $\Delta=2.4$ and $\Delta=2.8$ with moderate and strong effect modification. To estimate the sampling score, the combined trial and cohort sample is used to fit a weighted logistic model with $S_i$ as the outcome variable. To predict the potential outcomes in the combined trial and cohort data, linear models are fit among the trial sample. We simulate 5,000 data replications for each scenario and evaluate the performance of the two IPSW, REG and the two DR estimators. Both correct and incorrect model specifications are studied whenever applicable. In this simulation, a misspecified sampling score model does not include the interaction term $Z_{i1}Z_{i2}$, whereas an incorrect outcome model likewise omits $Z_{i1}Z_{i2}$. For each estimator, we consider the version that used the true treatment propensity score versus the version that used the estimated propensity score with $Z_{i1}$ and $Z_{i2}$. Such comparisons could illustrate the potential efficiency gain due to estimating the known propensity scores in the generalizability setting. Finally, the following quantities are computed for each scenario: the bias to $\Delta$, empirical standard error (ESE), average of the estimated standard errors (ASE), and empirical coverage probability of the 95$\%$ CIs constructed from the standard errors based on the proposed sandwich variance estimators. In addition to the main simulations, we also conduct additional simulations to assess the impact of misspecifying the target population size $N$ as well as a smaller trial sample size and/or cohort sample size. Those results are reported in Section \ref{sec:misspecify}. The \texttt{R} code for implementing the simulation studies can be found in Web Appendix 5.

\subsection{Main Simulation Results}\label{sec:sim2}
We report the simulation results with moderate selection effect and moderate effect modification in Table \ref{summary.table2}; results for the remaining three scenarios generally have similar patterns and are found in Web Tables 9 to 11. When the sampling score model is correctly specified, both IPSW1 and IPSW2 are unbiased and the associated 95\% CIs have close to nominal coverage with the use of proposed sandwich variance estimators. We also observe that using the estimated propensity scores could substantially reduce the variability of IPSW. Interestingly, although IPSW2 is more efficient than that IPSW1 in most scenarios, the former becomes less efficient than the latter when the both selection effect and effect modification become strong (Web Table 11). When the sampling score model is misspecified, IPSW is biased, even though the average sandwich standard error estimates stay close to the empirical standard errors; the bias of IPSW2 is larger than the bias of IPSW1 when the interaction term is omitted from the sampling score model.  When the outcome models are correctly specified, the REG estimator is consistent and more efficient than IPSW. As expected, when the outcome models are misspecified, the REG estimator has nontrivial bias. Weighting the outcome models by an estimated propensity scores has minimum effect on the efficiency of the REG estimator across all scenarios, which is concordant with the discussion in Section \ref{sec:asym3} that exploiting an estimated propensity score does not increase the asymptotic variance.

\begin{center}
[Table \ref{summary.table2} about here.]
\end{center}

The simulation results also demonstrate the robustness properties of the DR1 and DR2 estimators; that is, both estimators have negligible bias when either the sampling score model or the outcome models are correct, but not necessarily both. To further illustrate the double robustness and asymptotic normality, we present the empirical histograms of $\widehat{\Delta}_{\text{DR2}}$ (over 5000 simulations) under correct and incorrect model specifications in Web Figure 6. Furthermore, when all models are correctly specified, DR1 and DR2 are substantially more efficient than IPSW1 and IPSW2. For instance, when the true treatment propensity scores are used, the relative efficiency of $\widetilde{\Delta}_{\text{DR1}}$ to $\widetilde{\Delta}_{\text{IPSW1}}$ ranges from $1.46$ to $1.94$ while the relative efficiency of $\widetilde{\Delta}_{\text{DR2}}$ to $\widetilde{\Delta}_{\text{IPSW2}}$ ranges from $1.33$ to $2.02$. When the propensity scores are estimated from the trial sample, the relative efficiency of $\widehat{\Delta}_{\text{DR1}}$ to $\widehat{\Delta}_{\text{IPSW1}}$ ranges from $1.20$ to $1.69$ while the relative efficiency of $\widehat{\Delta}_{\text{DR2}}$ to $\widehat{\Delta}_{\text{IPSW2}}$ ranges from $1.21$ to $2.03$. Despite the efficiency gain over IPSW estimator, the DR estimators remain close but no more efficient than REG. Frequently, misspecification of the outcome models leads to more variable DR estimates than misspecification of the sampling score model, a phenomenon that is consistent with previous investigations in observational studies \citep{LiZaslavskyLandrum2013}. The average of the DR sandwich standard error estimates is generally close to the empirical standard error across all scenarios, indicating the adequacy of the proposed sandwich variance estimators. When all models are misspecified, DR1 and DR2 are biased for the true PATE.

Overall, DR1 and DR2 perform similarly across scenarios except when the outcome models are correctly specified but the sampling score model is not. In that case, DR2 shows higher efficiency over DR1, especially under strong selection effect for trial participation. For this reason, our simulation results favor DR2 over DR1. Correspondingly, using an estimated treatment propensity score generally has minimum effect on efficiency for both DR estimators, except when the sampling score model is correctly specified and the outcome models are not. In that case, estimating the known propensity score slightly improves the efficiency for both DR estimators under moderate effect modification in Table \ref{summary.table2}. Therefore, it may still be appealing to consider an estimated propensity score as it does not appear to adversely affect the finite-sample efficiency of DR estimators in the settings we considered.

\subsection{Additional Simulations}\label{sec:misspecify}

Our main simulation studies assume the target population size is known to be $N=$ 1 million, and relatively large compared to the trial and cohort sample sizes to mimic the two ACTG trials and the CNICS cohort. To generate additional empirical evidence, we conduct further assessments to investigate (1) the impact of misspecification of the target population size $N$, and (2) the performance of the generalizability estimators with a smaller trial sample size, $n$, and/or cohort sample size, $m$. We consider the most challenging scenario with a strong selection effect and strong effect modification with $\gamma=(-7.698,0.6,0.6,0.6)^T$ and $\zeta=(2,2,2)^T$. We only present the estimators with estimated propensity scores, as the results for the estimators with true propensity scores are completely analogous. Web Table 12 summarizes the performance of all five generalizability estimators when the target population size $N$ is underestimated to $0.8$ million and $0.5$ million (without altering the data generation process in Section \ref{sec:sim1}). Interestingly, the results for all five estimators are almost identical to those when the target population size is correctly specified. Similarly, the performance of all five estimators is also nearly unaffected when the target population size is overestimated to be $1.2$ million and $1.5$ million (Web Table 13). A likely explanation is that the true target population size $N$ is large enough such that there is a relatively large indifference range of $N$ within which the PATE estimates are relatively stable. This finding also matches our sensitivity analyses in Section \ref{sec:app} with slightly larger and smaller $N$, under which the PATE point and interval estimates remain nearly unchanged. Lastly, in the case when the target population size $N$ is severely underestimated to be $0.1$ million and $0.05$ million, Web Table 14 indicates that the bias of all estimators becomes nontrivial, with under-coverage especially in the latter scenario. However, with a correctly specified sampling score model even when $N=0.1$ million, the coverage of DR1 and DR2 estimators remains nominal, and  when $N=0.05$ million, both doubly-robust estimators have coverage over 90\% while both IPSW1 and IPSW2 can often exhibit notable under-coverage.

To investigate the performance of all five generalizability estimators with smaller (observed) sample sizes, we repeat the simulation study with $(n,m)\in\{(200,4000),(1000,800),(200,800)\}$, representing scenarios with a smaller trial sample size only, a smaller cohort sample size only, and smaller trial and cohort sample sizes. While the result patterns in Web Tables 15 and 16 are generally consistent with our main simulation with $(n,m)=(1000,4000)$, we observe that IPSW can exhibit excessive variance and under-coverage when the trial sample size decreases from $1000$ to $200$. In addition, IPSW2 appears to be less stable compared to IPSW1 when the trial sample size is small, demonstrating larger bias and lower coverage. In contrast, the performance of REG, DR1 and DR2 estimators are more stable when either the trial or the cohort sample size decreases. Even in the smallest sample size scenario with $(n,m)=(200,800)$, the DR estimators maintain nominal coverage when at least one model is correctly specified. Finally, while the accuracy of generalizability estimators is affected by both the trial sample size and the cohort sample size, the trial sample size appears to play a dominating role. This is expected because the trial sample contains information on both the effect modifiers and the outcome, whereas the cohort sample does not contain information on the outcome.

\section{Discussion}\label{sec:discussion}
In this article, we consider generalizing trial results from ACTG 320 and ACTG A5202 to two separate target populations. Our findings suggest the three-drug therapy with PI may lead to significant increase in CD4 cell counts among all people living with HIV, just as in ACTG 320. Likewise, the three-drug therapy with PI may lead to significant increase in CD4 cell counts among all women living with HIV, but with a potentially larger magnitude compared to women participants in ACTG 320. In contrast, the comparative evidence in ACTG A5202 appears less generalizable to the specified target populations. Unlike the compelling comparative evidence associated with ACTG 320, neither the SATE in ACTG A5202 nor the associated PATE estimate are significantly different from zero.

Our generalizability estimators require conditional exchangeability between the trial sample and population given measured covariates. This assumption is not testable without additional information of the outcome in the target population \citep{hartman2015sample} and could be violated if there exist unmeasured common causes of trial participation and the outcome. In cases where some potential effect modifiers are measured only in the trial but not in the target population, \citet{Nguyen2017} developed strategies for sensitivity analysis given assumed population-level information on the missing effect modifiers. It would be valuable for future work to adapt their approaches to our setting.

A second assumption we made is that the CNICS and WIHS cohorts are representative of the two target populations. Such an assumption is not directly testable with observed data and may be violated if participation in the cohort studies depends on demographic characteristics, access to health care, as well as medical history. To further address the difference between the cohort sample and target population, one needs to weight the cohort sample to approximate the covariate distribution of the target population. In the special case where the cohort study is a well-designed population survey with known survey weights, \citet{ackerman2021generalizing} developed the IPSW generalizability estimator that properly incorporated the survey weights. Using a similar strategy, it is possible to further extend our DR estimators in Section \ref{sec:estimation} by replacing $\Pi_0^{-1}$ with the known survey weights to estimate PATE.

While previous studies that estimate the PATE have implemented bootstrap approaches for inference \citep{Dahabreh2018a}, we have developed a set of closed-form sandwich variance estimators for inference. Our variance estimators extend the recent work of \citet{buchanan}, and are computationally more efficient than bootstrapping. Additionally, the development of the sandwich variance also has practical implications for our application because the we found certain interaction parameters in the outcome model are frequently not estimable during a bootstrapping procedure. For example, due to small trial sample size and sparse cell counts, there is no information to estimate the interaction between race and IDU, as well as IDU and baseline CD4 among 471 out of 1000 bootstrap replicates, when we generalize ACTG A5202 to all women living with HIV. This may raise concerns as one would have to change model specification for purpose of inference. The proposed sandwich variance estimator, to some extent, circumvents this issue, but still takes into account the uncertainty in estimating the parametric sampling scores, outcome models and/or propensity score model. In our simulations with comparable population sizes to the motivating application, the sandwich variance estimates are close to the empirical variances even under model misspecification. We provide \texttt{R} code for implementing the sandwich variance estimators in Web Appendix 5.

For estimating PATE, we formally demonstrated that generalizability estimators constructed with an estimated propensity score are asymptotically at least as efficient as those constructed with the true propensity score. In fact, using an estimated propensity score can be regarded as an implicit step to perform baseline covariate adjustment, which is known to increase the efficiency of within-trial SATE estimator \citep{Moore2011,shen2014inverse,williamson2014variance,Zeng2020}. Our simulation evidence suggests that using an estimated propensity score leads to more substantial efficiency gain for IPSW estimators and occasionally DR estimators with a misspecified outcome model. Our application also favors the use of a more saturated logistic model for estimating the propensity score in the trial, as it leads to substantially narrower CIs for PATE. While this strategy is supported by large-sample results, there is a tradeoff between asymptotic efficiency and finite-sample stability. In practice, when the trial is of a limited sample size (say less than 100), using a more saturated logistic model for the propensity score may result in overfitting, and can even compromise the finite-sample stability of the PATE estimates.

We acknowledge several limitations of our generalizability analyses which merit further study. First, we have created the cohort samples (CNICS and WIHS) by applying the ACTG trial inclusion criteria but without matching the recruitment years. Because the majority of our data is from the post-HAART era, we prioritized addressing the differences concerning the observed clinical characteristics rather than the unobserved differences related to calendar year. In other words, we have assumed that the trials and the harmonized cohorts are sampled from the same underlying target population, even if the recruitment years do not completely overlap. It would be of interest to ascertain additional data that also eliminate the temporal differences between trials and cohorts. Second, we have excluded the participants in ACTG trials with missing CD4 counts at follow-up. If the outcomes are not missing completely at random, the difference-in-means estimator of the SATE may be biased, which can lead to a biased IPSW estimate of PATE, even if the sampling score model is correctly specified. If the outcomes are missing at random, one could apply our generalizability estimators to multiply-imputed trial data sets and combine the results using the Rubin's rule. Finally, we have defined the causal estimand based on the entire population including the trial participants. An equally relevant estimand is the target average treatment effect (TATE) among the trial non-participants, defined as $\text{TATE} =\bE(Y^1-Y^0|S=0)$ \citep{Nguyen2017}. Estimation of TATE requires the use of inverse odds weights \citep{Westreich2017} instead of the inverse probability of sampling weights. When the cohort sample is a subset of the target population ($m<N$), it would be interesting to further investigate the necessity or potential benefit of incorporating the target population size $N$ to the inverse odds weights, similar to the arguments made in Section \ref{sec:prelim}.

\section*{Acknowledgements}
The authors thank the Editor, Associate Editor and the anonymous reviewer for their helpful comments, which greatly improve the exposition of this work. The authors also thank Issa J. Dahabreh, Michael G. Hudgens, Joseph J. Eron, Eric S. Daar, Michael J. Mugavero and Paul E. Sax for comments, and thank Can Meng for computational assistance with the simulation studies. Research reported in this publication was supported by the National Institute of Allergy and Infectious Diseases of the National Institutes of Health (NIH) under Award Number UM1 AI068634, UM1 AI068636 and UM1 AI106701. These findings are also presented on behalf of the Women's Interagency HIV Study (WIHS), the Center for AIDS Research (CFAR) Network of Integrated Clinical Trials (CNICS), and the AIDS Clinical Trials Group (ACTG). We would like to thank all of the WIHS, CNICS, and ACTG investigators, data management teams, and participants who contributed to this project. Funding for this study was provided by National Institutes of Health (NIH) grants U01AI042590, U01AI069918, 5-U01AI103390-02 (WIHS), R24AI067039 (CNICS), and P30AI50410 (UNC CFAR). Buchanan is partially supported by the Avenir Award Number 1DP2DA046856-01 from the National Institute on Drug Abuse of the NIH. The content is solely the responsibility of the authors and does not necessarily represent the official views of the NIH.

\bibliographystyle{jasa3}
\bibliography{DRGEN}

\clearpage

\begin{table}[htbp]
\caption{A synthesis of methods for generalizing trial results to target populations based on the potential outcomes framework, and classification by techniques for variance estimation, mode of inference and model assumptions.}
\vspace{0.1in}
\centering\label{tb:summary}
{\footnotesize
\begin{tabular}{lll}
\toprule
Design & Method & Reference \\
\midrule
Nested & Subclassification & \citet{stuart}$^{\diamond}$, \citet{tipton}$^{\dag}$, \citet{muirch}$^{\dag}$\\
& Weighting & \citet{stuart}$^{\diamond}$, \citet{Dahabreh2018a,Dahabreh2018c}$^{\natural}$\\
& Regression & \citet{Dahabreh2018a,Dahabreh2018c}$^{\natural}$\\
& Doubly robust & \citet{Dahabreh2018a,Dahabreh2018c}$^{\natural}$, \citet{li2021note}$^{\sharp}$\\
\midrule
Non- & Subclassification & \citet{buchanan}$^{\dag}$\\
nested & Weighting & \citet{cole10}$^{\dag}$, \citet{hartman2015sample}$^{\sharp}$, \citet{buchanan}$^{\ddag}$, this article$^{\ddag}$\\
& Regression & \citet{wang2019bayesian}$^{\S}$, this article$^{\ddag}$\\
& Doubly/multiply robust & \citet{rudolph2017robust}$^{\flat}$, \citet{lee2021improving}$^{\sharp}$, this article$^{\ddag}$\\
\bottomrule
\end{tabular}
\begin{tablenotes}
\item $\diamond$ variance estimation not discussed (\emph{frequentist, with parametric models});
\item $\natural$ implemented bootstrap variance (\emph{frequentist, with parametric models});
\item $\sharp$ implemented bootstrap variance (\emph{frequentist, calibration weighting});
\item $\dag$ implemented (sandwich) variance which ignores the uncertainty in estimating nuisance parameters (\emph{frequentist, with parametric models});
\item $\ddag$ implemented (sandwich) variance which takes into account the uncertainty in estimating nuisance parameters (\emph{frequentist, with parametric models});
\item $\S$ variance obtained from posterior summaries (\emph{Bayesian, with nonparametric priors});
\item $\flat$ variance estimated by sample variance of the efficient influence curve (\emph{frequentist, targeted maximum likelihood estimation with ensemble machine learner}).
\end{tablenotes}
}
\end{table}

\clearpage

\clearpage

\clearpage

\clearpage

\begin{table}
\caption{\footnotesize Estimated within-trial sample average treatment effect (SATE) and target population average treatment effects (PATE) on change in CD4 cell count$^a$ and corresponding 95$\%$ confidence intervals based on proposed sandwich variance estimators. Trial data are from ACTG 320$^b$ and ACTG A5202$^c$; the WIHS and CNICS cohorts are used to create random samples from two target populations (all people $N=1.1$ million and all women living with HIV in the USA $N = 280,000$). The three sets of results correspond to generalizability estimators that use (i) the true propensity score among trial participants $e_i=1/2$; (ii) the estimated propensity score using a logistic regression model with main effects for all covariates and a quadratic term for age; and (iii) the estimated propensity score using a more complex logistic regression model with main effects for all covariates, a quadratic term for age, and all pairwise interactions between covariates.}\label{tableall_k}
\vspace{0.1in}
		\footnotesize
	\hspace{-0.25in}\begin{tabular}{lccccccccc}
    \toprule
    & & & & SATE & PATE & PATE & PATE & PATE & PATE\\
     {Cohort}  &{$m$}  & {Trial } &{$n$} & & IPSW1 & IPSW2 & REG & DR1 & DR2\\
\midrule
 \multicolumn{6}{l}{\textit {True propensity scores}} & & & &  \\
		CNICS & 6,158 &   320 & 1,040 & 19 (12, 25)& 20 (11, 29)&18 (10, 26)&14 (5, 23)&15 (6, 24) & 15 (7, 24)\\
		 CNICS& 12,302 &  A5202& 1,440 & 6 (-8, 20)&5 (-27, 36)&1 (-26, 28)&-2 (-27, 24)&-3 (-29, 24)&-3 (-29, 24)\\	
     WIHS& 493  & 320  & 173 &24 (7,41)&52 (19, 86)&42 (15, 69)& 32 (-2, 65) &37 (1, 73)&37 (2, 72)\\
	   WIHS & 1,012 &  A5202  &  255 & 1 (-35, 37)  & 106 (-44, 256) & 31 (-44, 105)&0.04 (-103, 104)&4 (-84, 92)&4 (-85, 92)\\
		\midrule
		  \multicolumn{6}{l}{\textit{Estimated propensity scores with main-effects logistic model}} & & & &  \\
 		 CNICS & 6,158 &   320 & 1,040 &19 (12, 25)& 18 (10, 26)&17 (9, 25)&14 (5, 23)& 15 (7, 24)&15 (7, 24)\\
		 CNICS& 12,302 &  A5202& 1,440 &6 (-8, 20)&-2 (-26, 31)&0.33 (-24, 25)&-1  (-27, 25) &-1 (-27, 25)&-1 (-27, 25)\\
     WIHS& 493  & 320  & 173 & 24 (7,41)&41 (13, 69) &47 (21, 73)&35 (3, 68) &40 (6, 74)& 40 (6, 74)\\
	   WIHS & 1,012 &  A5202  & 255 &1 (-35, 37)  &81 (-49, 211) &37 (-37, 111)&10 (-90, 110)&11 (-75, 96)&11 (-75, 96)\\
		 \midrule
		  \multicolumn{6}{l}{\textit{Estimated propensity scores with full logistic model}} & & & &  \\
 		 CNICS & 6,158 &   320 & 1,040 &19 (12, 25)& 18 (10, 25)&18 (10, 25)&14 (5, 23)& 15 (6, 24)&15 (6, 24)\\
		 CNICS& 12,302 &  A5202& 1,440 &6 (-8, 20)&0.24 (-21, 22)&-1 (-24, 22)& -3 (-27, 23) &-4 (-29, 22)& -4 (-29, 22)\\
     WIHS& 493  & 320  & 173 & 24 (7,41)&38 (16, 60) &39 (17, 61)&25 (-6, 55) &28 (-3, 60)&28 (-3, 60)\\
	   WIHS & 1,012 &  A5202  & 255 &1 (-35, 37)  &52 (-39, 144)  & 24 (-28, 75)&-9 (-75, 57)&-8 (-75, 58)&-8 (-75, 59)\\
		 \bottomrule
  \end{tabular}
  \begin{tablenotes}
\item $^a$For ACTG 320, the outcome is change in CD4 cell count from baseline to week 4. For ACTG A5202, the outcome is change in CD4 cell count from baseline to week 48.
\item $^b$For ACTG 320, the treatment contrast is protease inhibitor ($X=1$) versus no protease inhibitor  ($X=0$).
\item $^c$For ACTG A5202, the treatment contrast is abacavir-lamivudine ($X=1$) versus tenofovir disoproxil fumarate-emtricitabine ($X=0$) plus efavirenz or ritonavir-boosted atazanavir.
\end{tablenotes}
\end{table}

\clearpage

\begin{table}[htbp]
\caption{Comparison of performance of five different estimators for estimating PATE with 5000 simulated data replications with $(\gamma_1,\gamma_2,\gamma_3)=(0.3,0.3,0.3)$ and $(\alpha_1,\alpha_2,\alpha_3)=(1,1,1)$ in the main simulation. The true PATE $\Delta=2.4$. ESE: Empirical standard error; ASE: Average of the estimated standard errors.}\label{summary.table2}
\vspace{0.1in}
\footnotesize
\renewcommand{\arraystretch}{0.8}
\centering
\begin{tabular}{lccrrrr}
\toprule
Estimator & Correct $w(Z_i;\gamma)$ & Correct $m_x(Z_i;\alpha_x)$ & Bias & ESE ($\times 100$)& ASE ($\times 100$)& Coverage ($\times 100$) \\
\midrule
\multicolumn{7}{l}{\emph{True treatment propensity scores used}} \\
$\widetilde{\Delta}_{\text{IPSW1}}$ & $\surd$ & -- & 0.00 & 12.3 & 12.9 & 95.7 \\
$\widetilde{\Delta}_{\text{IPSW1}}$ & $\times$ & -- & 0.02 & 12.8 & 13.5 & 95.3 \\
$\widetilde{\Delta}_{\text{IPSW2}}$ & $\surd$ & -- & 0.00 & 10.1 & 10.1 & 95.3 \\
$\widetilde{\Delta}_{\text{IPSW2}}$ & $\times$ & -- & -0.02 & 10.2 & 10.3 & 94.8 \\
$\widetilde{\Delta}_{\text{REG}}$ & -- & $\surd$ & 0.00 & 7.5 & 8.3 & 96.9 \\
$\widetilde{\Delta}_{\text{REG}}$ & -- & $\times$ & 0.03 & 8.1 & 8.9 & 95.9 \\
$\widetilde{\Delta}_{\text{DR1}}$ & $\surd$ & $\surd$ & 0.00 & 7.5 & 8.4 & 96.8 \\
$\widetilde{\Delta}_{\text{DR1}}$ & $\surd$ & $\times$ & 0.00 & 8.0 & 8.8 & 96.9 \\
$\widetilde{\Delta}_{\text{DR1}}$ & $\times$ & $\surd$ & 0.00 & 7.5 & 8.4 & 96.9 \\
$\widetilde{\Delta}_{\text{DR1}}$ & $\times$ & $\times$ & 0.09 & 8.2 & 9.1 & 85.3 \\
$\widetilde{\Delta}_{\text{DR2}}$ & $\surd$ & $\surd$ & 0.00 & 7.5 & 8.4 & 96.8 \\
$\widetilde{\Delta}_{\text{DR2}}$ & $\surd$ & $\times$ & 0.00 & 8.0 & 8.8 & 97.0 \\
$\widetilde{\Delta}_{\text{DR2}}$ & $\times$ & $\surd$ & 0.00 & 7.5 & 8.4 & 96.9 \\
$\widetilde{\Delta}_{\text{DR2}}$ & $\times$ & $\times$ & 0.09 & 8.2 & 9.0 & 85.2 \\
\midrule
\multicolumn{7}{l}{\emph{Estimated treatment propensity scores used}} \\
$\widehat{\Delta}_{\text{IPSW1}}$ & $\surd$ & -- & 0.00 & 9.0 & 9.7 & 96.7 \\
$\widehat{\Delta}_{\text{IPSW1}}$ & $\times$ & -- & 0.02 & 9.1 & 9.8 & 96.0 \\
$\widehat{\Delta}_{\text{IPSW2}}$ & $\surd$ & -- & 0.00 & 9.1 & 9.1 & 95.1 \\
$\widehat{\Delta}_{\text{IPSW2}}$ & $\times$ & -- & -0.02 & 9.1 & 9.1 & 94.7 \\
$\widehat{\Delta}_{\text{REG}}$ & -- & $\surd$ & 0.00 & 7.5 & 8.3 & 96.9 \\
$\widehat{\Delta}_{\text{REG}}$ & -- & $\times$ & 0.03 & 8.0 & 8.8 & 95.8 \\
$\widehat{\Delta}_{\text{DR1}}$ & $\surd$ & $\surd$ & 0.00 & 7.5 & 8.4 & 96.7 \\
$\widehat{\Delta}_{\text{DR1}}$ & $\surd$ & $\times$ & 0.00 & 7.9 & 8.8 & 96.9 \\
$\widehat{\Delta}_{\text{DR1}}$ & $\times$ & $\surd$ & 0.00 & 7.5 & 8.4 & 96.9 \\
$\widehat{\Delta}_{\text{DR1}}$ & $\times$ & $\times$ & 0.09 & 8.2 & 9.0 & 85.4 \\
$\widehat{\Delta}_{\text{DR2}}$ & $\surd$ & $\surd$ & 0.00 & 7.5 & 8.4 & 96.7 \\
$\widehat{\Delta}_{\text{DR2}}$ & $\surd$ & $\times$ & 0.00 & 7.9 & 8.8 & 97.0 \\
$\widehat{\Delta}_{\text{DR2}}$ & $\times$ & $\surd$ & 0.00 & 7.5 & 8.4 & 96.9 \\
$\widehat{\Delta}_{\text{DR2}}$ & $\times$ & $\times$ & 0.09 & 8.2 & 9.0 & 85.4 \\
\bottomrule
\end{tabular}
\end{table}

\end{document}